\documentclass[aps,prb,superscriptaddress,preprint]{revtex4}  
\usepackage{amssymb, amsmath} 
\usepackage{graphicx} 
\usepackage{makeidx} 

\newcommand{\ket}[1]{\mid \! #1 \rangle}

\begin{document}

\title{Theory of electronic properties and quantum spin blockade in a gated 
linear triple quantum dot with one electron spin each} 

\author{Chang-Yu Hsieh} 
\affiliation{Quantum Theory Group, 
Institute for Microstructural Sciences, 
National Research Council, Ottawa, Canada K1A 0R6} 
\affiliation{Department of Physics, 
University of Ottawa, Ottawa, ON, Canada, K1N 6N5} 

\author{Yun-Pil Shim} 
\affiliation{Quantum Theory Group, 
Institute for Microstructural Sciences, 
National Research Council, Ottawa, Canada K1A 0R6} 
\affiliation{Department of Physics, 
University of Wisconsin-Madison, Madison WI 53706} 

\author{Pawel Hawrylak} 
\affiliation{Quantum Theory Group, 
Institute for Microstructural Sciences, 
National Research Council, Ottawa, Canada K1A 0R6} 
\affiliation{Department of Physics, 
University of Ottawa, Ottawa, ON, Canada, K1N 6N5} 

\begin{abstract} 
We present a theory of electronic properties and the spin blockade phenomena in a gated linear triple quantum dot. 
Quadruple points where four different charge configurations are on resonance, particularly involving (1,1,1) configuration, are considered.
In the symmetric case, the central dot is biased to higher energy and a single electron tunnels through the device when (1,1,1) configuration is resonant with (1,0,1),(2,0,1),(1,0,2) configurations.
The electronic properties of a triple quantum dot are described by a Hubbard model containing two orbitals in the two unbiased dots and a single orbital in the biased dot. The transport through the triple quantum dot molecule involves both singly and doubly occupied configurations and necessitates the description of the (1,1,1) configuration beyond the Heisenberg model. Exact eigenstates of the triple quantum dot molecule with up to three electrons are used to compute current assuming weak coupling to the leads and non-equilibrium occupation of quantum molecule states obtained from the rate equation. The intra-molecular relaxation processes due to acoustic phonons and cotunneling with the leads are included, and are shown to play a crucial role in the spin blockade effect. We find a quantum interference-based spin blockade phenomenon at low source-drain bias and a distinct spin blockade due to a trap state at higher bias. We also show that, for an asymmetric quadruple point with (0,1,1),(1,1,1,),(0,2,1),(0,1,2) configurations on resonance, the spin blockade is analogous to the spin blockade in a double quantum dot. 

\end{abstract}

\maketitle

\section{Introduction} \label{sec:INTRO}
Gated quantum dots (QDs) \cite{ciorga_sachrajda_prb2000,hawrylak_prb1999, hanson_petta_rmp2007,koppens_buizert_nat2006,nowack_koppens_sci2007,pioroLadriere_obata_natphys2008,gaudreau_kam_apl2009,laird_taylor_prb2010} 
with controlled electron numbers are a testbed for probing fundamental many-body physics as well as a promising platform for building spintronics and quantum information processing (QIP) devices.\cite{korkusinski_hawrylak_wsr2007} 
Until recently, most experimental and theoretical investigations of quantum circuits based on electron spin focused on the single and double quantum dot (DQD) devices.\cite{hanson_petta_rmp2007,korkusinski_hawrylak_wsr2007}. 
Many essential tasks for operating a qubit have been demonstrated in DQDs. For instance, coherent manipulation and readout of one\cite{koppens_buizert_nat2006}  and two\cite{petta_johnson_sci2005} spin states have already been experimentally achieved using spin blockade. \cite{ciorga_sachrajda_prb2000,johnson_petta_prb2005,ono_austing_sci2002} 
In DQDs, spin blockade is used to detect spin using spin-to-charge conversion. For instance, the $(0,2)$ charge configuration cannot be obtained from the $(1,1)$ configuration if the electron spin in the left dot is parallel to the electron spin in the right dot. Detected charge on the right dot depends on the relative spin orientations of the two electrons. Thus, spin blockade detects spin states (triplet or singlet) of the two electrons in transport spectroscopy or charge sensing measurement. \cite{ciorga_sachrajda_prb2000,johnson_petta_prb2005,ono_austing_sci2002} A physical signature of spin blockade at the triple point, $(0,1) \rightarrow (1,1) \rightarrow (0,2)$, is the current rectification under different bias directions.  In positive (forward) bias direction, triplet states will not be populated, and the system does not manifest negative differential conductance. In negative (reverse) bias direction, current suppression is pronounced once  the transitions to the $(1,1)$ triplet states become accessible in the transport window.

A nontrivial extension of the quantum circuit based on electron spin is the triple quantum dot (TQD) with one electron each. This can be appreciated by the comparison of the quantum optical properties of a two-level versus three-level systems. Charging and transport spectroscopy experiments
\cite{gaudreau_kam_apl2009,granger_gaudreau_prb2010,korkusinski_gimenez_prb2007,groveRasmussen_jorgensen_nanolett2008}
on the TQDs have already mapped out the stability diagram of the devices down to a few electrons. 
Recent experiments\cite{gaudreau_granger_arxiv2011, laird_taylor_prb2010} have also demonstrated coherent manipulations of 
electron spins in TQDs.
The electronic properties of a TQD have been investigated theoretically, including topological Hunds rules,\cite{korkusinski_gimenez_prb2007} 
spin-selective Aharonov-Bohm oscillations,\cite{shim_delgado_prb2009,delgado_shim_prl2008}  
the implementation of a coded qubit,\cite{diVincenzo_bacon_nat2000,hawrylak_korkusinski_ssc2005,hsieh_hawrylak_prb2010} 
voltage-controlled spin manipulation,\cite{shim_sharma_ssc2010,shim_hawrylak_prb2008}
entangled GHZ state generation,\cite{rothlisberger_lehmann_prl2008,sharma_hawrylak_prb2011} 
non-Fermi-liquid behaviour \cite{ingersent_ludwig_prl2005,zitko_bonca_prl2007,lobos_aligia_prb2006} 
in a triangular TQD as well as coherent tunneling adiabatic passage (CTAP) processes for a single electron in a linear triple quantum dot (LTQD). \cite{greentree_cole_prb2004,emary_prb2007}   
All these theoretical predictions as well as quantum information processing in a TQD require an ability to spectroscopically detect spin by, e.g., spin blockade.

In recent experiments Granger\cite{granger_gaudreau_prb2010} {\it et. al.} and Laird \cite{laird_taylor_prb2010} {\it et. al.} carried out transport spectroscopy and charge sensing measurement on a LTQD molecule with one electron in each dots. This configuration, denoted by $(1,1,1)$, was tuned to be resonant with the two electron configuration $(1,0,1)$. It was assumed that transport proceeded through $\{(2,0,1),(1,1,1),(1,0,2)\}$ resonant configurations, which implied that the central dot was biased to higher energy. The presence of doubly occupied dots in the configurations makes the Heisenberg model of localized spin configurations inapplicable and a microscopic model is required to study the electronic and transport properties of this TQD system.

Here we extend our earlier theory of a TQD \cite{hawrylak_korkusinski_ssc2005,korkusinski_gimenez_prb2007,gimenez_korkusinski_prb2007,gimenez_hsieh_prb2009}
to biased linear molecule at quadruple points (QPs) and describe spin blockade as a spectroscopic tool allowing the readout of electron spin.  
We analyze the electronic and spin properties of a LTQD as a function of energies of each dot within a single-band or multi-band Hubbard model. 
The knowledge of the wave functions of a single-band Hubbard model allows for the qualitative understanding of the low-bias transport through the device, but including more than one orbital in the dot will be shown to be crucial for spin blockade. 
Two different QPs involving the $(1,1,1)$ configuration are considered: (a) symmetrical QP (SQP) with $(1,0,1),(2,0,1),(1,1,1),(1,0,2)$ configurations on resonance,
and (b) asymmetrical QP (AQP) with $(0,1,1),(1,1,1),(0,2,1),(0,1,2)$ configurations on resonance. For SQP, the transport goes through $(1,0,1)\rightarrow (2,0,1) \rightarrow (1,1,1) \rightarrow (1,0,2)$ channels, while $(0,1,1) \rightarrow (1,1,1) \rightarrow (0,2,1) \rightarrow (0,1,2)$ is the transport channel for the AQP.
Current is calculated in sequential tunneling approximation between the TQD and the leads, using rate equations\cite{shim_delgado_prb2009,muralidharan_datta_prb2007} to calculate the non-equilibrium steady state occupation of TQD states with a source-drain bias. We use Fermi's Golden Rule to calculate the transition rates between TQD states by adding or removing an electron due to the coupling between the TQD molecule and the leads, and also the transition rate between TQD states with the same number of electrons due to the interaction with acoustic phonons.\cite{florescu_hawrylak_prb2006,inarrea_platero_prb2008}

The plan of the paper is as follows. In Sec.\ref{sec:model} and Sec.\ref{sec:elecprop}, we describe the system, the Hamiltonian, and the electronic properties of a TQD as a function of detuning $\Delta$ of the central dot.  In Sec.\ref{sec:seqtunnel}, our approach to the transport based on the sequential tunneling between the leads and the TQD molecule and rate equations are explained in detail. The transition rates due to different mechanisms are also discussed.  In Sec.\ref{sec:SQP}, we present results of current calculations for the SQP, and discuss the mechanism of quantum spin blockade at low bias.  In Sec.\ref{sec:AQP}, we present results of transport calculations for conventional spin blockade at the SQP under high source drain bias and at the AQP, and discuss how the system at the AQP can behave qualitatively as a double dot around a similar triple point with $(0,1), (1,1), (0,2)$ configurations. A brief conclusion is given in Sec.\ref{sec:CON}.

\section{Model} \label{sec:model}
Figure \ref{fig:layout} presents a schematic diagram of a LTQD in contact with the two semi-infinite leads and the energy levels of the single QD orbitals. The metallic leads are modelled by one dimensional tight binding chains. Each quantum dot, defined by metallic gates on top of GaAlAs/GaAs heterojunction and represented here by a circle contains a controlled number of electrons, e.g., one electron each [$(1,1,1)$ configuration] in (a) and $(1,0,1)$ configuration in (b). Electrons can tunnel between dots $1$ and $2$, and between dots $2$ and $3$, but there is no direct tunnel coupling between the two edge dots. Figure \ref{fig:layout}(c) shows the single particle levels of the individual dots in the LTQD without interdot tunneling. The lowest energy bars denote $S$ orbitals (the ground orbitals) in each dot. The energy of the central dot is raised by an applied voltage $\Delta$. This bias can be used, for example, in order to localize the two electrons in dots $1$ and $3$ as shown in Fig. \ref{fig:layout}(b).  In this study, $\Delta$, comparable to Coulomb repulsion $U$, is used to bring the configurations such as $(1, 1, 1)$ and $(1, 0, 2)$ on resonance as shown in Fig. \ref{fig:enres}. 
We find it is essential to include the excited states, $P$ orbitals, in dots $1$ and $3$ in order to properly account for the transport properties at the SQP.  The energy separation, $\Delta_{sp}$, between $S$ and $P$ orbitals may also be comparable to $\Delta$.  Thus, the electronic properties of a LTQD are described by a multi-band Hubbard model with parameters derived from  a microscopic Linear Combination of Harmonic Orbitals-Configuration Interaction (LCHO-CI) approach for given voltages on the gates. \cite{gimenez_korkusinski_prb2007}
With $\hat{c}_{i\sigma}$ ($\hat{c}^{\dag}_{i\sigma}$) denoting annihilation (creation) operators for an electron with spin $\sigma$ on orbital $i$, the  five-level Hubbard Hamiltonian reads:
\begin{equation}
\label{eq:hubbard}
\hat{H}_{D} = \sum_{i=1, \sigma}^{5} E_{i}(V_{sd}) \hat{n}_{i\sigma} + \sum_{\substack{i,j=1,\sigma \\ j \neq i}}^{5}t_{ij}\hat{c}^{\dag}_{i\sigma}\hat{c}_{j\sigma} + 
\sum_{i=1}^{5}  U_{i} \hat{n}_{i \uparrow} \hat{n}_{i\downarrow} +\frac{1}{2} \sum_{i,j=1 }^{5} V_{ij} \hat{\rho}_{i} \hat{\rho}_{j},
\end{equation}
where $E_{i}(V_{sd})$ is the sorce-drain bias dependent energy of orbital $i$, and $t_{ij}$, $U_{i}$, and $V_{ij}$ are tunnel coupling, on-site and off-site Coulomb repulsion between orbitals $i$ and $j$ respectively,  
$\hat{n}_{i\sigma} = \hat{c}^{\dag}_{i\sigma} \hat{c}_{i \sigma}$, and $\hat{\rho}_{i} =\sum_{\sigma} \hat{n}_{i\sigma}$.  
We assign indices $i=1,2,3$ to $S$ orbitals of dots $1$, $2$ and $3$ respectively.  The indices $i=4, 5$ denote excited $P$ orbitals. We will consider only single excited orbitals in both dot 1 ($i$=4) and dot 3 ($i$=5) for the TQD molecule at SQP. For the AQP case with $(0,1,1)$ base configuration, the excited orbitals are in dot 2 ($i$=4) and dot 3 ($i$=5). The excited orbital in the biased dot does not play any significant role.

The TQD device is connected to left and right leads ($r = L, R$)  as shown in Fig. \ref{fig:layout}(a). Electrons in the leads fill up the noninteracting states of semi-infinite tight-binding chains with a bulk dispersion relation $\epsilon_r(k) = 2t_r \cos\left(ka\right)$ up to a Fermi level $\mu_{L(R)}$, 
where $t_r$ is the tunnel coupling between the sites on lead $r$, $a$ is the distance between sites of tight-binding chain, and $k$ denotes the mode of the plane wave for single particle states in the chain. The interaction between the leads and the device is modelled as,
\begin{equation}
\label{eq:lead-dot}
\hat{H}_{rD} = \sum_{i_{r}, \sigma} \sum_{k} \left( \tilde{t}_{i}^r (k) \hat{d}^{\dag}_{k\sigma} \hat{c}_{i_{r} \sigma} + h.c. \right),
\end{equation}
where $\tilde{t}_{i}^r (k) = t_{i}^{r}e^{i2 \pi ka m_r}/\sqrt{2\pi}$ is the tunnel coupling between the mode $k$ of the $r=L (R)$ lead and orbital $i_r$ localized in the left dot ($r=L$) or the right dot ($r=R$) and $m_r$ in the exponent of $\tilde{t}_{i}^r(k)$ is $1$ for $r=L$ and $-1$ for $r=R$. $\hat{d}^{\dag}_{k\sigma}$ creates an electron with momentum $k$ and spin $\sigma$ in the lead $r$.  In this study, $t_i^R=0$ for orbitals not in the right edge dot and $t_i^L=0$ for orbitals not in the left edge dot. 

Interactions with phonons have already been shown to be important to understand the incoherent transport properties of  double quantum dots at high bias in Ref.~\onlinecite{inarrea_platero_prb2008}, for instance. We include interaction of electrons in the LTQD with bulk longitudinal acoustic (LA) phonons via deformation potential as the mechanism of phonon-induced relaxation at low temperature. The electron-phonon interaction Hamiltonian reads,
\begin{equation}
\label{eq:ep}
\hat{H}_{e-ph} = \sum_{i,j=1,\sigma}^{5}\sum_{q} M_{ij}(\mathbf{q}) \left( \hat{b}_{\mathbf{q}} + \hat{b}^{\dag}_{-\mathbf{q}}\right) \hat{c}^{\dag}_{i\sigma}\hat{c}_{j\sigma},
\end{equation}
where $\mathbf{q}$ is the phonon momentum, $i$ and $j$ are TQD orbitals, and $\hat{b}_{\mathbf{q}} (\hat{b}^{\dag}_{\mathbf{q}})$ operator annihilates (creates) a phonon with momentum $\mathbf{q}$.   $M_{ij}(\mathbf{q}) = \Lambda(q) \int \psi_i(r)^* \exp(-i\mathbf{q} \cdot \mathbf{r}) \psi_j(r)$ is the electron-phonon scattering matrix element,  $\psi_i(r)$ is a single particle wave function, and $\Lambda(q) =  \sqrt{\frac{D^2 \hbar q}{2 \rho c_s}}$ for deformation potential $D$, GaAs mass density $\rho$, and speed of sound $c_s$ in GaAs.  The phonon scattering matrix element, $M_{ij}(\mathbf{q})$, depends on the  single particle wave function $\psi_i(r)$ which is obtained from the LCHO \cite{gimenez_korkusinski_prb2007} formalism.

\subsection{Electronic Properties of a LTQD} \label{sec:elecprop}

The electronic properties of a triangular TQD molecule with all three dots on resonance for $N=1-6$ electrons have been described in detail in Ref.\onlinecite{korkusinski_gimenez_prb2007}. We focus here on the linear molecule where there is no tunneling between the end quantum dots and on the effect of detuning $\Delta$ of the energy of the central dot. 
While in numerical calculations we retain all five levels, we retain only the three lowest energy states in this semi-analytical discussion of the low energy spectra. 
For the Hubbard parameters, we set $U_{i} = U$, $t_{12}= t_{23} = t$, $t_{13}=0$, $V_{13} = V$, and $V_{12} = V_{23} = V^{'}$.  For the on-site energies, we restrict our attention to  $E_{1} = E_{3} =E$ and $E_{2} = E+\Delta$.  Since $E$ is just an overall shift in energy, we will simply set $E=0$ until we explicitly state otherwise.  The Hubbard Hamiltonian commutes with total $\hat{S}^2$ and $\hat{S}_y$ , so we consider spin-resolved subspaces in the Hilbert space.

First, we focus on the single particle molecular states of the TQD. We consider the $S_y = 1/2$ subspace and use a localized basis $\{ \vert 1 \rangle, \vert 2 \rangle, \vert 3 \rangle \}$ , where $\vert i \rangle = c^{\dag}_{i\uparrow}\vert 0 \rangle$.  In this basis, the Hubbard Hamiltonian, Eq.(\ref{eq:hubbard}), reads, 
\begin{equation}
\label{eq:Hubb1e}
H_{1e} = \left[ \begin{array}{ccc}
0 & t & 0 \\
t & \Delta & t \\
0 & t & 0 \\
\end{array} \right].
\end{equation}
By inspection we see that a state $\vert D \rangle = (\vert 1 \rangle - \vert 3 \rangle)/\sqrt{2}$,  with an energy $E_{D}=0$, is an eigenstate. In this state, an electron does not occupy the central dot. This state can block the transport in a setting where dots 1 and 3 connected to the source and dot 2 connected to the drain, and hence is called a dark state, \cite{michaelis_emary_epl2006,emary_prb2007} in analogy to the coherent population trapping in quantum optics.
The existence of a dark state can be detected by transport spectroscopy\cite{michaelis_emary_epl2006} of an empty dot. As the transport window determined by the applied source-drain voltage $V_{sd}$ is large enough to allow the added electron to enter a dark state, a negative differential conductance should be observed in the experiment.  Furthermore, Greentree\cite{greentree_hamilton_prb2004,greentree_cole_prb2004} {\it et. al.} proposed to implement CTAP to move an electron from dot one to dot three without passing through dot two and for quantum information transfer for a double-dot charge qubit.

There are two states orthogonal to the dark state $\vert D \rangle$: the bright state $\vert B  \rangle = (\vert 1 \rangle + \vert 3 \rangle)/\sqrt{2}$ and the central state $\vert C \rangle = \vert 2 \rangle$ .  The 2-by-2 Hamiltonian matrix spanned by the bright and central states can be analytically diagonalized, and the two eigenstates are expressed as a linear combination of the bright and central state: 
$\vert M_1 \rangle = \cos(\phi) \vert B \rangle + \sin(\phi) \vert C \rangle$ and
$\vert M_2 \rangle = -\sin(\phi) \vert B \rangle + \cos(\phi) \vert C \rangle$,
where $\tan(2\phi) = -\sqrt{2}t/\Delta$.
We note that tuning $\phi$ allows us to recover Jacobi eigenstates discussed in Ref.\onlinecite{korkusinski_gimenez_prb2007,gimenez_hsieh_prb2009, hsieh_hawrylak_prb2010}. 
Tuning $t$ mostly controls the amount of mixing between the bright and central state in the two eigenstate $\vert M_1 \rangle$ and $\vert M_2 \rangle$, whereas tuning $\Delta$ can control the energy spacing between the two eigenstates.  The energies associated with the three eigenstates $\vert D \rangle, \vert M_1\rangle,$ and $ \vert M_2 \rangle$, are  $E_{D}=0$, $E_{M_{1}}= (\Delta - \Delta_t)/2$, and $E_{M_{2}}= (\Delta + \Delta_t)/2$ where $\Delta_t = \sqrt{\Delta^2 + 8t^2}$.  We note that $\vert M_1 \rangle$ is always the ground state.

Next, we address the two-electron case. The simpler case to analyze is the triplet $S_y = 1$ subspace, which contains two spins up in the LTQD.  There are 3 basis vectors $ \{ \vert T_{1} \rangle$, $\vert T_{2} \rangle$, $\vert T_{3}\rangle \}$, where $\vert T_{1} \rangle = \hat{c}^{\dag}_{2\uparrow}\hat{c}^{\dag}_{1\uparrow} \vert 0 \rangle$, $\vert T_{2} \rangle = \hat{c}^{\dag}_{3\uparrow}\hat{c}^{\dag}_{1\uparrow} \vert 0 \rangle$,  and $\vert T_{3} \rangle = \hat{c}^{\dag}_{3\uparrow}\hat{c}^{\dag}_{2\uparrow} \vert 0 \rangle$, respectively.   The Hubbard Hamiltonian in this basis reads
\begin{equation}
\label{eq:Hubb2el_1}
H_{2T} = \left[ \begin{array}{ccc}
\Delta+V^{'} & t & 0 \\
t & V  & t \\
0 & t & \Delta+V^{'} \\
\end{array} \right].
\end{equation}
The triplet Hamiltonian, Eq.(\ref{eq:Hubb2el_1}), and the single particle Hamiltonian, Eq.(\ref{eq:Hubb1e}), have the identical matrix structure.  Therefore, there is a dark triplet eigenstate 
$\vert T_D \rangle = (\vert T_{1} \rangle - \vert T_{3} \rangle )/\sqrt{2} $ and the bright $\vert T_B \rangle = (\vert T_{1}  \rangle + \vert T_{3} \rangle)/\sqrt{2}$ and central state, $\vert T_C \rangle = \vert T_{2} \rangle$.  Rotating the Hamiltonian, Eq. (\ref{eq:Hubb2el_1}) into the basis of bright, central, and dark states,  a 2-by-2 Hamiltonian matrix coupling the bright and central states is derived.  The two eigenstates of the triplet subspace are 
 $\vert M^T_1 \rangle = \sin(\phi) \vert T_B  \rangle + \cos(\phi) \vert T_C \rangle$ and 
 $\vert M^T_2 \rangle = -\cos(\phi) \vert T_B \rangle + \sin(\phi) \vert T_C \rangle$, 
where $\tan(2\phi) = \sqrt{2}t/\Delta_v $, and $\Delta_v = \Delta + V^{'} -V$ . The corresponding eigenenergies of the three states are
\begin{subequations}
\begin{align}
\label{eq:Hubb_tripen}
 E_{T_{D}} & =   \Delta + V^{'} , \\
 E_{M^{T}_{1}} & =   \Delta + V^{'} - \frac{1}{2}\left( \Delta_v+\sqrt{(\Delta_v)^2+8t^2} \right)  , \\
 E_{M^{T}_{2}} & =   \Delta + V^{'} - \frac{1}{2} \left( \Delta_v-\sqrt{(\Delta_v)^2+8t^2} \right).
\end{align}
\end{subequations}
The ground state $\vert M^T_1 \rangle$ is predominantly characterized by $\vert T_C \rangle = \vert \uparrow_1 \uparrow_3 \rangle$ with spins up in dots $1$ and $3$ because the corresponding coefficient $\sin(2\phi) \approx  1- \frac{t^2}{2\Delta_v^2}$ when $t/\Delta$ is small.  Nevertheless, $\vert M^T_1 \rangle$ still has non-zero presence in both $\vert T_{1} \rangle = \vert \uparrow_1 \uparrow_2 \rangle$ and $\vert T_{3} \rangle = \vert \uparrow_2 \uparrow_3 \rangle$ configurations.
In later sections, we will explain how the low bias spin blockade formation is related to the small yet finite components of $\vert T_1 \rangle$ and $\vert T_3 \rangle$ in $\vert M^T_1 \rangle$ wave function.  
We designate three ground states in each of the spin-resolved triplet subspaces with $(S=1,S_y=1,0,-1)$ as $\ket{T^{+}}$, $\ket{T^0}$, and $\ket{T^{-}}$, respectively.
$\ket{T^{+}}$ = $\ket{M^T_1}$ as was shown above, and $\ket{T^0}$ is obtained by flipping one spin and performing symmetrization of the wavefunction 
and $\ket{T^{-}}$ is obtained by flipping both spins from $\ket{T^{+}}$. These states will play the major roles in the transport through LTQD at the low source-drain bias
Furthermore, we find it also useful to represent $\ket{T^{+}}$ as:
\begin{equation}
\label{eq:sb_triplet}
\vert T^+ \rangle = \gamma \left( \ket{T_{2}} + \frac{\gamma_1}{\gamma} \ket{T_{1}}+\frac{\gamma_2}{\gamma}\ket{T_{3}}\right),
\end{equation}
where the coefficients $\vert \gamma_{1(2)} \vert \ll 1$.

Next, we analyze the $S_y = 0$ singlet state for 2 electrons.  We define the following basis $\{ \vert S_{1} \rangle, \vert S_{2}\rangle, \vert S_{3}\rangle, \vert S_{4} \rangle, \vert S_{5} \rangle, \vert S_{6} \rangle \}$.  The singly occupied configurations are, 
$\vert S_{1} \rangle = \frac{1}{\sqrt{2}} \left( \hat{c}^{\dag}_{1\downarrow} \hat{c}^{\dag}_{2\uparrow} +  \hat{c}^{\dag}_{2\downarrow} \hat{c}^{\dag}_{1\uparrow} \right) \vert 0 \rangle$, 
$\vert S_{2} \rangle = \frac{1}{\sqrt{2}} \left( \hat{c}^{\dag}_{1\downarrow} \hat{c}^{\dag}_{3\uparrow} +  \hat{c}^{\dag}_{3\downarrow} \hat{c}^{\dag}_{1\uparrow} \right) \vert 0 \rangle$, and 
$\vert S_{3} \rangle = \frac{1}{\sqrt{2}} \left( \hat{c}^{\dag}_{2\downarrow} \hat{c}^{\dag}_{3\uparrow} +  \hat{c}^{\dag}_{3\downarrow} \hat{c}^{\dag}_{2\uparrow} \right) \vert 0 \rangle$.
The doubly occupied configurations are,
$\vert S_{4} \rangle = \hat{c}^{\dag}_{1\downarrow} \hat{c}^{\dag}_{1\uparrow} \vert 0 \rangle$, 
$\vert S_{5} \rangle = \hat{c}^{\dag}_{2\downarrow} \hat{c}^{\dag}_{2\uparrow} \vert 0 \rangle$, and 
$\vert S_{6} \rangle = \hat{c}^{\dag}_{3\downarrow} \hat{c}^{\dag}_{3\uparrow} \vert 0 \rangle$.  
The Hubbard Hamiltonian in this basis reads, 
\begin{equation}
\label{eq:Hubb2el_2}
H_{2S} = \left[ \begin{array} {cccccc}
\Delta+ V^{'} & t & 0 & \sqrt{2}t & \sqrt{2}t  & 0   \\
t & V &t & 0 & 0 & 0 \\
0 & t & \Delta+V^{'} & 0 & \sqrt{2}t & \sqrt{2} t \\
\sqrt{2} t &  0  & 0 &  U & 0 & 0 \\
\sqrt{2} t  & 0 & \sqrt{2} t  &  0 & 2\Delta +U & 0 \\
0 & 0 & \sqrt{2}t & 0 & 0  &   U 
\end{array}\right].
\end{equation}
The 3-by-3 upper left block, spanned by $ \{ \vert S_{1} \rangle, \vert S_{2}\rangle, \vert S_{3}\rangle \}$, 
is identical to the triplet Hamiltonian, Eq. (\ref{eq:Hubb2el_1}). 
For $\vert \Delta \vert$ small compared to on-site Coulomb repulsion $U$, the energy spectrum of the singlet subspace can be divided into the bands of singly occupied and doubly occupied configurations, with a gap of the order of $U$. Under such condition, the energies and wavefunctions of the first three lowest singlet states are very similar to those of the triplet states, and the mixing between singly and doubly occupied configurations leads to a $t-J$ model.\cite{shim_hawrylak_prb2008}  
However, if $\vert \Delta \vert $ is comparable to $U$, then the singlet subspace has a ground state predominantly characterized by $\vert S_{2} \rangle$ configuration, which is well separated from the four excited states characterized by $\vert S_{1} \rangle, \vert S_{3}\rangle, \vert S_{4} \rangle, \vert S_{6} \rangle$.  
The doubly occupied $\vert S_{4} \rangle$ and the singly occupied $\vert S_{1} \rangle$ configurations, which are connected by tunneling between dot 1 and dot 2, get very close in energy. When these states are degenerate, two eigenstates can be obtained by $\vert U_{14}^{\pm} \rangle = \frac{1}{\sqrt{2}}(\vert S_1 \rangle \pm \vert S_4 \rangle)$. Similarly, we get $\vert U_{36}^{\pm} \rangle = \frac{1}{\sqrt{2}}(\vert S_3 \rangle \pm \vert S_6 \rangle)$ from $\vert S_{3} \rangle$ and $\vert S_{6} \rangle$.
By second order perturbation theory, the well-isolated ground state with dominant contribution from $\vert S_2 \rangle$, has energy
 \begin{equation}
 \label{eq:Hubb_singen}
 E_{M^{S}_1} = V-4t^2\left( \frac{1}{\Delta+V'+U-V+2\sqrt{t}} +  \frac{1}{\Delta+V'+U-V-2\sqrt{t}} \right).
  \end{equation}
We note that the singlet-triplet splitting  is $E_{M^{T}_{1}} - E_{M^{S}_{1}} > 0$ for all range of  $\Delta_v$ and $t$, and we have singlet as the ground state.

Next, we consider the three electron states.  
In the fully spin polarized subspace, $S_y= + 3/2 $, there is only one  state 
$\vert S=3/2,S_y=+3/2 \rangle =\hat{c}^{\dag}_{3\uparrow} \hat{c}^{\dag}_{2\uparrow} \hat{c}^{\dag}_{1\uparrow} \vert 0 \rangle$, 
with energy  given by $E^{3/2} = 3E+\Delta+2V^{'}+V$.  
This state is characterized by having a spin up electron in each dot. 
For $S_y = + 1/2$ subspace, it is composed of 9 singly and doubly occupied configurations. 
To simplify the qualitative analysis, we focus on a truncated basis composed of the following 3 singly occupied configurations: 
$ \vert a \rangle = \hat{c}^{\dag}_{3\uparrow} \hat{c}^{\dag}_{2\uparrow} \hat{c}^{\dag}_{1\downarrow} \vert 0 \rangle$, 
$\vert b \rangle = \hat{c}^{\dag}_{3\uparrow} \hat{c}^{\dag}_{2\downarrow} \hat{c}^{\dag}_{1\uparrow}\vert 0 \rangle,$
$\vert c \rangle = \hat{c}^{\dag}_{3\downarrow} \hat{c}^{\dag}_{2\uparrow} \hat{c}^{\dag}_{1\uparrow}\vert 0 \rangle,$
and two doubly occupied configurations 
$\vert d \rangle =  \hat{c}^{\dag}_{3\uparrow} \hat{c}^{\dag}_{1\uparrow} \hat{c}^{\dag}_{1\downarrow}\vert 0\rangle$, 
$ \vert e \rangle = \hat{c}^{\dag}_{3\uparrow} \hat{c}^{\dag}_{3\downarrow} \hat{c}^{\dag}_{1\uparrow}\vert 0 \rangle$.
Figure \ref{fig:enres} shows resonance between configuration $\vert b \rangle $ and $\vert e \rangle$
when $\vert \Delta \vert = O(U)$.  
The three-dimensional subspace with singly occupied configurations with $S_y$=1/2 can be further decomposed by the total spin $S$, since $S$ is also a good quantum number. For the subspace with $S$=1/2, we use the Jacobi basis states $ L_0$ and $L_1$:\cite{diVincenzo_bacon_nat2000,hawrylak_korkusinski_ssc2005}  
$\vert L_0 \rangle = \frac{1}{\sqrt{2}}( \vert a \rangle - \vert c \rangle)$, 
$\vert L_1 \rangle = \frac{1}{\sqrt{6}} (\vert a \rangle - 2\vert b \rangle + \vert c \rangle)$. 
For $\vert L_0 \rangle$, the spin state in dot 1 and dot 3 is a singlet.  For $\vert L_1 \rangle$,  the spin state in dots 1 and dot 3  can be written as a linear combination of triplets with $S_y = 0$ and $S_y=1$.  
The remaining Jacobi state 
$\vert L_2 \rangle = \frac{1}{\sqrt{3}} ( \vert a \rangle + \vert b \rangle + \vert c \rangle) $
is a total spin $3/2$ state and is decoupled from all other states.
In a similar fashion, we form Jacobi coordinates for the two doubly occupied configurations
$ \vert X \rangle = \frac{1}{\sqrt{2}} (\vert d \rangle + \vert e \rangle)$, 
and $\vert Y \rangle = \frac{1}{\sqrt{2}} (\vert d \rangle -\vert e \rangle)$. 
In the subspace of $S$=1/2 and $S_y$=1/2, with basis $\{ \vert L_0 \rangle,\vert X  \rangle, \vert L_1 \rangle, \vert Y \rangle \}$, the 3-electron Hamiltonian,
 \begin{equation}
 \label{eq:Hubb3el}
 H_{3el} = \left[ \begin{array} {cccc}
 \Delta+2V^{'}+V & -t & 0 & 0 \\
-t & U + 2V & 0 & 0 \\
0 & 0 & \Delta+2V^{'} + V & \sqrt{3} t \\
0 & 0 &  \sqrt{3}t & U + 2V
 \end{array} \right],
 \end{equation}
separates into the pair of Hamiltonians describing Jacobi basis states $\vert L_0 \rangle$ and $\vert L_1 \rangle $ entangled with the doubly occupied configurations.  Each sub-matrix can be diagonalized and the eigenstates read:
 $ \vert L_0^{+} \rangle = \cos(\phi) \vert L_0 \rangle + \sin(\phi) \vert X \rangle$,
 $ \vert L_0^{-} \rangle = \sin(\phi) \vert L_0 \rangle  -   \cos(\phi) \vert X \rangle$,
 $ \vert L_1^{+} \rangle = \cos(\theta) \vert L_1 \rangle + \sin(\theta) \vert Y \rangle$,
and
 $ \vert L_1^{-} \rangle = \sin(\theta) \vert L_1 \rangle - \cos(\theta) \vert Y \rangle$,
where $\tan(2\phi) = t/\xi$, $\tan(2\theta) = \sqrt{3}t/\xi$, and $\xi = (\Delta+2V'-U-V )/2$.
Here, we observe that each of the two Jacobi states, characterizing the $(1,1,1)$ configuration, hybridizes with both doubly occupied configurations $|X\rangle$ and $|Y\rangle$ to form the eigenstates of a central-dot biased system.  All four eigenstates, $\vert L_0^{\pm} \rangle$ and $\vert L_1^{\pm} \rangle$, are current-conducting because electrons can be removed from the orbitals in the edge dots to make a transition from the three-electron state to a two electron $(1,0,1)$ configuration.


In Fig. {\ref{fig:sqp_cdoten}(a) we  show the evolution of the five lowest energy levels of the three electron complex in the $S_y=1/2$ subspace as a function of bias $\Delta$ in the central dot.  At $\Delta=0$, the spectrum is divided into 2 bands.  The lower band consists of $\vert L_1 \rangle$, $\vert L_0 \rangle$, and $\vert L_2 \rangle$ states, which are all characterized by singly occupied configurations.  The upper band consists of states with dominant configurations $\vert d \rangle$ and $\vert e \rangle$. As $\Delta$ increases, the energy difference between the singly occupied configurations and specific doubly occupied configurations $\vert d \rangle$ and $\vert e \rangle$ diminishes.  However, the ground state is always the $\vert L^+_1\rangle$ state in the figure. The blue curve represents the spin-3/2 state which does not interact with all other levels due to the conservation of total spin of the Hamiltonian.  In the plot, the levels are artificially shifted for better visualization.  In the inset of Fig. \ref{fig:sqp_cdoten}(a), the proper energy levels around the anti-crossing point are shown in detail.  
Figure \ref{fig:sqp_cdoten}(b) shows the configuration content of the ground state as a function of bias $\Delta$. At $\Delta=0$ the ground state is dominated by singly occupied configuration $\vert L_1 \rangle$ but at higher bias, $\Delta \approx U$, the doubly occupied configuration $\vert Y \rangle$ reaches around $50\%$  content of the ground state.
 
\subsection{Current through a linear triple quantum dot} \label{sec:seqtunnel}
Theory of sequential tunneling through a triangular TQD molecule has been described in detail in Ref. \onlinecite{shim_delgado_prb2009}. Here we extend the approach to include both electron-phonon interaction and cotunneling and apply this theory to describe current and spin blockade in a LTQD. Following Ref. \onlinecite{shim_delgado_prb2009}, current between lead $r$ and a TQD device in the vicinity of a QP involving $N=2$ and $N+1=3$ electrons can be written as a difference between the current from the lead to the TQD and a current from the TQD back to the lead $r$:
\begin{align}
\label{eq:curr}
I_{rD} & = -e \sum_{i_r, \sigma}\sum_{\alpha_N, \beta_{N+1}} W^{seq}_{r}(\alpha_N \rightarrow \beta_{N+1}) P_{\alpha_{N}} \nonumber \\
 & + e \sum_{i_r, \sigma} \sum_{\alpha_{N}, \beta_{N+1}} W^{seq}_{r}(\beta_{N+1} \rightarrow \alpha_{N}) P_{\beta_{N+1}},
\end{align}
where
$\vert \alpha_N \rangle$ is an $N$-electron many-body eigenstate of the isolated TQD with energy $E_{\alpha_{N}}$ and  associated steady state probability $P_{\alpha_N}$, which is obtained by solving the rate equation, which is explained below. The sequential tunneling rate,  $W^{seq}_r(\alpha_N \rightarrow \beta_{N+1})$, provides the rate of transition for the TQD from an N-electron $\alpha_N$ state to an ($N+1$)-electron state due to first order perturbation from the lead $r$.  Details of sequential tunneling rates will be provided later.

The probabilities $P_{\alpha_{N}}$'s are the diagonal matrix elements of the reduced density matrix $\rho$.  The time evolution of these diagonal matrix elements is described by the Pauli master equation,
\begin{equation}
\label{eq:rate}
\dot{P}_{\alpha_N} = \sum_{N'=2}^{3} \sum_{\beta_{N'}} P_{\beta_{N'}} W({\beta_{N'}\rightarrow \alpha_{N}})  -  P_{\alpha_{N}} W({\alpha_{N}\rightarrow \beta_{N'}} ),
\end{equation}
where transition rates $W_{\alpha_{N} \rightarrow \beta_{N'}}$ are calculated using Fermi's Golden Rule.  We consider sequential tunneling rate $W^{seq}_r$ in first order in coupling to the lead $r$, intra TQD phonon-induced relaxation rate $W^{ph}$, and second order cotunneling rate $W^{cot}_r$.
The master equation is solved to obtain steady-state solution for the probabilities, $P_{\alpha_{N}}$, by setting the time derivatives to be zero.

With the coupling to a lead $r$ in Eq. (\ref{eq:lead-dot}), the first order sequential tunneling rates read
\begin{subequations}
\begin{align}
\label{eq:tunnelrate}
W_r^{seq}(\alpha_N \rightarrow \beta_{N+1}) & =  \frac{2\pi}{\hbar}\sum_k \left|  \langle \beta_{N+1} \vert  \sum_{i} \tilde{ t}_{i}^{r}(k) \hat{c}^{\dag}_{i\sigma} \vert \alpha_{N} \rangle \right| ^2  \delta \left(\omega_{\alpha\beta}-\epsilon_{rk}\right) f_r(\omega_{\alpha \beta}), \\
W_r^{seq}(\beta_{N+1} \rightarrow \alpha_{N}) & = \frac{2\pi}{\hbar}\sum_k  \left| \langle \alpha_{N} \vert \sum_{i}\tilde{t}_{i}^{r}(k)  \hat{c}_{i \sigma} \vert  \beta_{N+1} \rangle \right| ^2 \delta \left(\omega_{\alpha\beta}-\epsilon_{rk}\right) (1-f_r(\omega_{\alpha\beta})), 
\end{align}
\end{subequations}
where $f_r(\epsilon) = 1/\left(\exp[(\epsilon-\mu_r)/k_BT]+1\right)$ is the Fermi function of the lead $r$, $\omega_{\alpha \beta} = E_{\beta_{N+1}}-E_{\alpha_N}$, and  $\epsilon_{rk}$ is the energy of a state associated with wave vector $k$ of lead $r$. We remark that the summation over index $i$ in the sequential tunneling rate refers to summing the tunneling contributions from the $S$ and $P$ orbitals in a quantum dot.
By expanding the norms of the complex-valued matrix elements in above equations and introducing an integration variable $\omega$, the sequential tunneling rates
can be also expressed as follows,
\begin{subequations}
\label{eq:spectralfnseq}
\begin{align}
W_r^{seq}(\alpha_N \rightarrow \beta_{N+1}) & = \frac{2\pi}{\hbar} \sum_{i, j} \int d\omega A^{\alpha \beta}_{ij} (\omega) B^r_{ij}(\omega) f_r(\omega_{\alpha \beta}) , \\
W_r^{seq}(\beta_{N+1} \rightarrow \alpha_{N}) & = \frac{2\pi}{\hbar} \sum_{i,j} \int d\omega  A^{\alpha \beta}_{ij} (\omega) B^r_{ij}(\omega) (1-f_r(\omega_{\alpha\beta})),
\end{align}
\end{subequations}
with generalized spectral functions \cite{vaz_kyriakidis_jcp2009} of the TQD, $A^{\alpha \beta}_{ij} = \sum_{\sigma} \langle \alpha_N \vert \hat{c}_{i\sigma} \vert \beta_{N+1} \rangle \langle \beta_{N+1} \vert \hat{c}^{\dag}_{j\sigma} \vert \alpha_N \rangle \delta(\omega-\omega_{\alpha\beta})$, and generalized spectral function of the lead $r$, $B^r_{ij}(\omega) = \sum_{k} \tilde{t}^r_i(k)(\tilde{t}^r_j(k))^* \delta(\omega-\epsilon_{rk})$.  By substituting the sequential tunneling rates in Eq.(\ref{eq:curr}) with Eq.(\ref{eq:spectralfnseq}), one can relate the current through a TQD with the spectral functions of the TQD and the leads. 

We now provide relaxation rates due to  electron-phonon interaction and cotunneling.
For large source-drain bias voltage   $\vert eV_{sd} \vert \gg |t_{ij}|$, 
the change in the on-site energy of dots due to the source-drain bias will take the system off the resonance, away from the QP. In this regime, the current is 
dominated by inelastic tunneling between orbitals of neighbouring quantum dots due to electron-phonon interaction.  The phonon emission-induced relaxation rate \cite{inarrea_platero_prb2008} reads,
\begin{equation}
\label{eq:phononrate}
W^{ph}\left(\alpha_N \rightarrow \beta_N\right) = \frac{2\pi}{\hbar} \sum_{\mathbf{q}} \left| \sum_{i,j,\sigma} M_{ij}(\mathbf{q})
\langle \beta_N \vert \hat{c}_{i\sigma} \hat{c}^{\dag}_{j\sigma} \vert \alpha_N \rangle\right|^2
 \delta \left(E_{\alpha_N}-E_{\beta_N}-\hbar\omega_{\mathbf{q}} \right) g(\hbar\omega_{\mathbf{q}},T),
\end{equation}
where $\hbar \omega_{\mathbf{q}} = \hbar c_s \vert \mathbf{q} \vert$ is phonon energy, and $g(\hbar\omega_{\mathbf{q}})$ is the thermal occupation number for phonon mode $\mathbf{q}$ at temperature $T$.  
Spin blockade occurs when the spin-3/2 polarized states $\ket{\alpha_3}$ become a trap state, with $W_r^{seq}(\alpha_3 \rightarrow \beta_2) = 0$. However, the spin blockade can be lifted if we allow cotunneling. We consider cotunneling transition rate,\cite{hansen_mujica_nanolett2008, qassemi_coish_prl2009} which involves an exchange of electrons between a lead $r$ and the TQD in a spin 3/2 state, 
\begin{align}
\label{eq:cotrate}
W^{cot}_r \left(\alpha_3 \rightarrow \beta_3\right) & =  & \frac{2\pi}{\hbar}\sum_{\sigma, \sigma^{'}, k, k'} F_r(\epsilon_{k'_{r}}^{\sigma'})(1-F_r(\epsilon_{k_{r}}^\sigma))
\delta \left( E_{\beta_{3}}-E_{\alpha_{3}}-\epsilon_{k_{r}}^\sigma-\epsilon_{k'_{r}}^{\sigma'} \right)  \nonumber \\
& & \times
 \left| \sum_{\gamma_2 , i, i'} t_{i}^r(k_r) t_{i'}^r(k'_{r'})\frac{\left(C^{i'_r\sigma^{'}}_{\alpha_3 \gamma_2}\right)^* C^{i_r\sigma}_{\beta_3\gamma_2}}{E_{\alpha_{3}}-E_{\gamma_{2}}-\epsilon_{k'_{r}}^{\sigma'}} \right|^2, 
\end{align}
where $C^{i\sigma}_{\alpha_3, \gamma_{2}} =  \langle \alpha_3 \vert \hat{c}^{\dag}_{i\sigma} \vert \gamma_2 \rangle$, $\vert \gamma_2 \rangle$ is a triplet state, and $\epsilon_{k_r}^\sigma$ is the energy for an electron with wave vector $k$ and spin $\sigma$ of the lead $r$.

\section{Transport and Spin Blockade} \label{sec:Res}
In this section, we compute and discuss the transport properties and spin blockade in a LTQD at both SQP and AQP. 
We set on-site Coulomb repulsion between $S$ orbitals to be $U_{11} = U_{22} = U_{33} = U =3.0$ meV, and we use $U$ as the unit of energy scale.  For $S$ and $P$ orbitals in the same dot, we set $U_{14} = U_{35} = U' = 0.94 U$, and $U_{44} = U_{55} = U'' = 0.96 U$ for $P$ orbitals in the same dot.  We set $t_{i,j} = t = -6.0\cdot10^{-3}\,U$  for tunneling between $S$ orbitals in neighbouring dots.  We set $t_{{ij}}= t' = -6.2\cdot10^{-3}\, U$ for tunneling between the $S$ and $P$ orbitals on neighbouring dots.  
We set $V_{ij} = V'=0.2\,U$ between neighbouring dots and $V_{ij} = V=0.1\,U$ between dots $1$ and $3$.  The energy difference between $S$ and $P$ energy levels, $\Delta_{sp}$, in the same dot is taken to be $0.8$ U and $0.25$ U in different cases considered below. 

The tunnel coupling for the tight binding chain in the leads is taken as $t_{L} = t_R = -2.0 U$.  The large tunnel coupling for the leads allows a wide energy band, which increases the amount of available states for transport.  As for the dot-lead tunnel coupling $t_i^r$, we set $ t_{1}^L =  -1.0\cdot10^{-3} U$ and $t_{4}^L = -1.1\cdot10^{-3} U$.  Only the $S$ and $P$ orbital in dot 1 is connected to the left lead.  Symmetrically, we set $t_{3}^R = t_{1}^L$, and $t_{5}^R = t_{4}^L$.  The rest of the tunnel coupling parameters are zero in our model.
For interaction between electrons in the TQD  and bulk LA phonons, we use the following GaAs parameters: $\Lambda(q) = \sqrt{D^2\hbar\omega_{q}/2\rho c_s^2}$, where $D = 2.9\, U$, $\rho = 5300\, kg / m^3$, $c_s = 3700 m/s$ and $\omega_{q} = c_s q$.

We measure current in unit of $I_0 = e \vert t_{1}^L \vert ^2 / {\hbar \vert t_L \vert}$.  We assume total potential difference  $eV_{sd}$ across the two leads and  a linear decrease of this potential across the device.  The chemical potentials on the two leads are given by $\mu_L = eV_{sd}/2$ and $\mu_R = -eV_{sd}/2$. The on-site energies are given by $E_{1,(4)}(V_{sd}) = E^0_{1,(4)}+eV_{sd}/6$,
$E_{i_{2}} = E^0_{i_{2}}$, and $E_{3, (5)}(V_{sd}) = E^0_{3,(5)}-eV_{sd}/6$ respectively, and electron temperature in all calculations is set to $k_BT = 2.0\cdot10^{-3}\, U$.

\subsection{Quantum Interference-Based Spin Blockade} \label{sec:SQP}
We first consider transport through the SQP: $\{(1,0,1)$, $(2,0,1)$, $(1,1,1)$, $(1,0,2)\}$, and we put $P$ orbitals ($i=4,5$) in dot 1 and 3, respectively.   We set $E_1 = E_3 = -U-V$ and $E_2 =-2V'$ in order to bring the four charge configurations into resonance. For the present case, we set a high single particle level spacing $\Delta_{SP} = 0.8U$ in the edge dots.  A large energy spacing between the $S$ and $P$ orbitals allows one to focus on a few lowest states for transport at bias $ \vert eV_{sd} \vert \ll  U$.  For instance, Fig. \ref{fig:sqp_en} shows the energy diagrams of the relevant 2-electron and 3-electron states near the SQP as a function of $V_{sd}$.   In the presence of small $V_{sd}$, the energy spectrum does not alter much and the wavefunctions remain similar to the wavefunctions at zero $V_{sd}$.  The inset in Fig. \ref{fig:sqp_en} categorizes the states associated with the energies in the main figure.  There are four active 2-electron states: one singlet $\ket{S}$, and triply degenerate triplet states, $\ket{T^{\pm,0}}$.  These four states are characterized predominantly by the $(1,0,1)$ charge configurations as discussed in Sec.\ref{sec:elecprop}.
For $N=3$ subspace, there are four spin-1/2 states below the spin-3/2 states.  The four spin-1/2 states are  $\vert L^+_1\rangle$ and $\vert L^+_0 \rangle$ and their counterpart in the $S_y=-1/2$ subspace.  In the absence of magnetic field, these states remain degenerate.  Next up in the  three-electron subspace are the quadruply degenerate spin-3/2 states, $\ket{S=3/2, S_y=\pm 1/2,\pm 3/2}$.  The last four levels are $\vert L^-_1 \rangle$ and $\vert L^-_0\rangle$ states and their counterpart in $S_y = -1/2$ subspace. We emphasize that these three electron states are admixtures of $(2,0,1)$, $(1,1,1)$, and $(1,0,2)$ configurations with comparable weights except the spin-polarized states as discussed in Sec.\ref{sec:elecprop}.  Based on the analysis of wavefunctions, obtained from the exact diagonalization of a single-band Hubbard Hamiltonian, the only dark channels in the LTQD are the spin-3/2 states.  As the spin-3/2 wavefunctions, $\ket{S=3/2,S_y=\pm1/2, \pm3/2}$, do not overlap significantly with the two-electron triplet states, $\ket{T^{\pm,0}}$, when an electron is added or removed from the edge dots, the conventional spin blockade is not expected in this regime. 

Figure \ref{fig:qmblockade}(a) shows the current $I(V_{sd})$  of a LTQD and Fig. \ref{fig:qmblockade}(b) shows the steady state occupation probability of the four spin-3/2 states as functions of $V_{sd}$. This was done without the cotunneling effect. 
The $I-V_{sd}$ curve is symmetrical with respect to the bias direction as it should be at SQP. The most prominent feature is that
the vanishing of the current and therefore significant negative differential conductance associated with high occupation probability of the spin-3/2 states.  
As shown in Fig. \ref{fig:qmblockade}(a), the current is completely suppressed at a very limited bias regime, this is very different from the $I-V$ curve in the spin blockade regime in a DQD.  These numerical results are obtained from a five-level Hubbard model, and the negative differential conductance is not reproduced when we use just the three-level Hubbard model for the transport calculation. This implies that this negative differential conductance is related to the existence of the high-energy $P$ orbitals.

In order to explain this negative differential conductance, we need to study $S_y=3/2$ subspace with all five orbitals. There are 10 possible configurations for three spin-up electrons in five orbitals.  Using the Hubbard model with these five orbitals, the configuration with the lowest energy is  $\ket{\bar{a}}= \hat{c}^{\dag}_{1\uparrow} \hat{c}^{\dag}_{2\uparrow} \hat{c}^{\dag}_{3\uparrow} \vert 0 \rangle$, and the next two configurations are $\ket{\bar{b}} =  \hat{c}^{\dag}_{1\uparrow} \hat{c}^{\dag}_{3\uparrow} \hat{c}^{\dag}_{5\uparrow} \vert 0 \rangle$ and $\ket{\bar{c}} =  \hat{c}^{\dag}_{1\uparrow} \hat{c}^{\dag}_{3\uparrow} \hat{c}^{\dag}_{4\uparrow} \vert 0 \rangle$.  Configuration $\ket{\bar{a}}$ is separated from $\ket{\bar{b}}$ and $\ket{\bar{c}}$ by an energy gap of $\sim U+\Delta_{sp}-\Delta$.  The other 7 configurations are even further away in energy.  The Hamiltonian of this low energy configuration subspace in the basis of $\{ \ket{\bar{b}}, \ket{\bar{a}}, \ket{\bar{c}} \}$ is,
\begin{equation}
\label{eq:3el2levelHmtx}
H_{3/2} = \left[ \begin{array}{ccc}
E_1+2E_3+\Delta_{sp}+U'+2V & -t' & 0 \\
-t' & E_1 + E_2 + E_3 + 2V' + V & -t' \\
0 & -t' & 2E_1+E_3+\Delta_{sp}+U'+2V
\end{array} \right],
\end{equation}
where $E_1$ and $ E_3$ are almost identical when $V_{sd}$ is small.  This Hamiltonian matrix looks similar to the 2-electron triplet Hamiltonian, Eq.(\ref{eq:Hubb2el_1}), except that the tunneling matrix elements acquire a negative sign for the three electron system.  This negative sign is simply due to the anticommutation relation between fermionic operators.  Exact diagonalization of the above Hamiltonian gives a ground state,
\begin{equation}
\label{eq:sb_polarized35}
 \vert 3/2 \rangle = \eta \left(\ket{\bar{a}} 
   + \frac{\eta_1}{\eta}\ket{\bar{b}} + \frac{\eta_2}{\eta}\ket{\bar{c}} \right) ,
\end{equation}
where coefficients $\eta_{1(2)}$ are of the same order of magnitude as the coefficients $\gamma_{1(2)}$ for $\vert T^+ \rangle$ in Eq.(\ref{eq:sb_triplet}).  This can be understood by analyzing the Hamiltonians.  The energy difference between the configuration $\ket{T_{1}}$ and $\ket{T_{3}}$ is given by  $\vert \Delta +V'-V \vert$, and the energy difference between the configurations $\ket{\bar{b}}$ and $\ket{\bar{a}}$ is given by $\vert \Delta-\Delta_{sp}-U'-V+2V'\vert$.   Considering that $\Delta$ and $\Delta_{sp}$ are both of the order of $U$, the two energy gaps are actually comparable. In general, hybridization of configurations $\ket{i}$ and $\ket{j}$ in a wavefunction can be estimated by $\frac{\langle i \vert H \vert j \rangle}{E_i - E_j}$.  In our case, the $S$-$P$ tunnel coupling $t'$ is of the same order of magnitude as the $S$-$S$ tunnel coupling $t$.  This explains why $\eta_{1(2)}$ are comparable to $\gamma_{1(2)}$ in magnitude.  Furthermore, $\eta_{1(2)}$ have opposite signs with respect to $\gamma_{1(2)}$ because the off-diagonal matrix elements in Eq.(\ref{eq:3el2levelHmtx}) and Eq.(\ref{eq:Hubb2el_1}) have opposite signs.  Figure \ref{fig:qmblockade}(c) presents the norm of $ \gamma_1$ and $\eta_1$ from the exact diagonalization of the five-level Hubbard model as a function of $V_{sd}$. 

Next, we look at the rate equation for the state $\vert 3/2 \rangle$ when the system is subject to a positive source-drain bias,
i.e., charging electron from left dot and removing electron from right dot:
\begin{align}
\label{eq:ratepolarized}
\frac{dP_{3/2}}{dt} = -W^{seq}_R(3/2 \rightarrow T^+) P_{3/2}+W^{seq}_L(T^+ \rightarrow 3/2) P_{T^+}~. 
\end{align}
Note that the only allowed 2-electron state is $\vert T^+ \rangle$ because the total spin cannot change by more than $1/2$ by adding an electron.  The phonon relaxation does not play a role here because $\vert 3/2 \rangle$ and $\vert T^+ \rangle$ are the lowest energy states in their own spin-resolved subspaces, respectively. 
For simplicity, we have ignored the cotunneling contribution in this analysis, and numerical results in Fig. \ref{fig:qmblockade} are also obtained without the cotunneling terms.
Cotunneling effects will be discussed below.  
In order for $\vert 3/2 \rangle$ to be a trap state, the outgoing part of the rate equation should be almost equal to zero.  
The outgoing sequential rate is, approximately,
\begin{align}
\label{eq:polarizedseqrate}
W^{seq}_R (3/2 \rightarrow T^+)  & =   \frac{2\pi}{\hbar}\sum_k \left \vert \langle T^+  \vert  \tilde{t}^R_3(k)  \hat{c}_{3\uparrow} \vert 3/2 \rangle 
 + \langle T^+ \vert  \tilde{t}^R_5(k) \hat{c}_{5\uparrow} \vert 3/2 \rangle \right \vert ^2 \nonumber \\
&\times \delta(E_{3/2}-E_{T^{+}}-\epsilon_{kR}) (1-f_R), \nonumber \\ 
& = \frac{2\pi}{\hbar}\sum_k  \left \vert \left(\vert  \gamma_1 \vert t^R_3 -\vert \eta_1 \vert t^R_5 \right) \frac{e^{-ika}}{\sqrt{2\pi}} \right \vert ^2  \delta(E_{3/2}-E_{T^{+}}-\epsilon_{kR}) (1-f_R),
\end{align}
where $\tilde{t}^R_i(k) = t^R_i e^{-ika}/\sqrt{2\pi}$ and $f_R$ is the fermi function for the right lead. The coefficients $\gamma_1$ and $\eta_1$ are defined in Eq.(\ref{eq:sb_triplet}) and Eq.(\ref{eq:sb_polarized35}), respectively.  The expression  $\left( \vert \gamma_1 \vert t^R_3 -\vert \eta_1 \vert t^R_5 \right)$ gives the interference between the two possible paths of removing an electron (via the $S$ and $P$ orbital) from the right dot.  The minus sign in the expression stems from the fact that $\eta_{1}$ and $\gamma_{1}$ have opposite signs, and the origin of this sign difference was already explained immediately following Eq.(\ref{eq:3el2levelHmtx}). We see that the condition for the quenching of the sequential tunneling rate is  $\vert \gamma_1 / \eta_1 \vert = \vert t^R_5 / t^R_3 \vert$.  Figure \ref{fig:qmblockade}(d) presents the ratio $\vert \gamma_1 / \eta_1 \vert$ as a function of $V_{sd}$. At points of strongest current suppression, we observe that the ratio indeed matches the ratio of $\vert t^R_5 / t^R_3 \vert$.  In short, the negative differential conductance sets in whenever the two possible paths of electronic transport become comparable in amplitude and interfere destructively.  This destructive interference is possible only for the transport channels through spin-3/2 states. In terms of spin configurations, the transport channel $\vert 3/2 \rangle \rightarrow \vert T^+ \rangle$ involves the two paths $(\uparrow_{1},\uparrow_{2},\uparrow_{3}) \rightarrow (\uparrow_{1},\uparrow_{2})$ and $(\uparrow_{1},\uparrow_{3},\uparrow_{5}) \rightarrow (\uparrow_{1},\uparrow_{3})$, which can destructively interfere. 
For all other transport channels, electronic transport occurs with much higher probability amplitude via the $S$ orbital in the edge dots at low source-drain bias.  
The existence of the dark channel through $\vert 3/2 \rangle$ makes the TQD molecule to be trapped in $\vert 3/2 \rangle$ state.

Figures \ref{fig:qmblockade2}(a) and (b) show the current through the LTQD and the steady state probability distribution for the spin-3/2 states in the parameter space of $(E_1=E_3, E_2)$ at a small bias, $eV_{sd} = 0.01$ $U$, respectively .  In this calculation, the cotunneling effect is included. Although the quantum interference-based spin blockade is formed under a very specific condition, Fig. \ref{fig:qmblockade2}(b) shows that the interference-based spin blockade can still be observed in the parameter space of on-site energies.  The cotunneling effects can be analyzed when we add terms $\sum_{r=L,R} \sum_{\beta_3} W^{cot}_r(\beta_3 \rightarrow 3/2)P^{\beta_3}-\sum_{r=L,R}\sum_{\beta_3} W^{cot}_r(3/2 \rightarrow \beta_3)P^{3/2} $, where $\ket{\beta_3}$ is a three-electron state with $S=1/2$, in Eq.(\ref{eq:ratepolarized}).  In Eq.(\ref{eq:polarizedseqrate}), we analyze the condition for the transition rate from $\ket{3/2}$ to $\ket{T^+}$ to vanish.  With cotunneling included in the model, we should analyze the condition for the transition rate from $\ket{3/2}$ to each $\ket{\beta_3}$ state to vanish.  In principle, each transport channel has its unique condition for the quenching, and the interference-based quantum spin blockade will be lifted.  However, the additional rates due to cotunneling are much smaller in amplitude as they scale with $\vert t^L_1 \vert^4$ for the second order processes. The system still gets blockaded in the spin-3/2 state when the sequential tunneling driven transition ($W^{seq}_R(3/2 \rightarrow T^+)$) vanishes, because the incoming rate $W^{seq}_L(T^+ \rightarrow 3/2)$, a first order process, scales with $\vert t^L_1 \vert^2$ and is around 5 orders of magnitude larger than the rates driven by cotunneling processes. 

\subsection{Symmetrical and Asymmetrical Spin Blockade} \label{sec:AQP}

Next, we look at the SQP again with a different single-particle level spacing, $\Delta_{sp} = 0.25$ U.  In this case, we will consider a wider range of source-drain bias with $eV_{sd} > U$.
Figure \ref{fig:sqp}(a) shows the current, $I(V_{sd})$, of the TQD near the SQP. We again have a symmetric $I(V_{sd})$ with respect to the bias directions and, therefore, the observed negative differential conductance is also bi-directional. We will focus on the positive bias direction for the following discussion. We note that there are 2 regions where the current is strongly suppressed in the positive bias direction in Fig. \ref{fig:sqp}(a).  One point is at the low bias regime, $eV_{sd} \ll U$, and the other  point is at the high bias regime such that on-site triplet occupation is allowed in the transport window.  From Fig. \ref{fig:sqp}(b), we see that the system is trapped in $(1,1,1)$ spin 3/2 states whenever the current is significantly suppressed in Fig. \ref{fig:sqp}(a). The strong current suppression at the low bias is due to the quantum interference-based spin blockade we described in the previous section. As source-drain bias is further increased, the wavefunction inside the LTQD also changes. Gradually one path of electronic transport becomes preferred and quantum interference vanishes. At the high bias, the second current suppression is identified to be the more familiar spin blockade phenomenon in the double quantum dot, and it is characterized by an extended region of current suppression over a wider range of source-drain bias. At high bias, hybridization of levels becomes insignificant, and it is instructive to look at each eigenstate as a particular localized configuration.  Figure \ref{fig:phononblockade} presents a schematics of how this high bias spin blockade is formed and lifted in the LTQD at high bias.  The spin blockade is formed when the on-site triplet becomes accessible in the left dot but not in the right dot in the transport window when a positive bias is applied.  Due to the phonon-induced relaxation, the on-site triplet in the left dot will relax by allowing electron-phonon scattering to re-distribute the electron from the $P$ orbital in the edge dot onto the $S$ orbital in the central dot.  When the on-site triplet state in the right dot is still too high in energy for occupation, the system gets stuck in this $(1,1,1)$ spin-3/2 configuration. This spin blockade is lifted when the bias is further increased so the on-site triplet become accessible in the right dot too. Then the phonon-induced relaxation will again help transfer the electron from the central dot onto the right dot. We remark that the spin blockade does not happen in this model if the phonon-induced relaxation mechanism is removed.  
From this picture, we can derive the spin blockade regime from the parameters we used. The energies of the relevant configurations are
\begin{eqnarray}
E(\uparrow_1 \uparrow_3) &=& E_1 + E_3 + V = -2.1 U ~, \nonumber\\
E(\uparrow_1 \uparrow_3 \uparrow_4) &=& 2 E_1 + \Delta_{sp} + E_3 + U' + 2 V + \frac{eV_{sd}}{6} = -1.91U + \frac{eV_{sd}}{6} ~,\nonumber\\
E(\uparrow_1 \uparrow_2 \uparrow_3) &=& E_1 + E_2 + E_3 + 2 V + V' = -2.1 U ~, \nonumber\\
E(\uparrow_1 \uparrow_3 \uparrow_5) &=& E_1 + 2 E_3 + \Delta_{sp} + U' + 2 V - \frac{eV_{sd}}{6} = -1.91U - \frac{eV_{sd}}{6} ~.\nonumber 
\end{eqnarray}
For an electron to move from the left lead to the TQD, $E(\uparrow_1 \uparrow_3)+\epsilon_e$=$E(\uparrow_1 \uparrow_3 \uparrow_4)$ for an electron energy $\varepsilon_e \le \mu_L = eV_{sd}/2$. Thus, we get $ eV_{sd} \ge 0.57U$. $E(\uparrow_1 \uparrow_2 \uparrow_3)$ is always lower than $E(\uparrow_1 \uparrow_3 \uparrow_4)$ for forward bias, so the relaxation from $|\uparrow_1 \uparrow_3 \uparrow_4 \rangle$ to $|\uparrow_1 \uparrow_2 \uparrow_3 \rangle$ is allowed. For the spin blockade to occur, the transition from $|\uparrow_1 \uparrow_2 \uparrow_3 \rangle$ to $|\uparrow_1 \uparrow_3 \uparrow_4 \rangle$ should not be possible by phonon emission. So, $E(\uparrow_1 \uparrow_2 \uparrow_3) < E(\uparrow_1 \uparrow_3 \uparrow_5)$, which leads to $eV_{sd} < 1.14U$. Therefore, the spin blockade regime is $0.57U \le eV_{sd} < 1.14U$, which agrees very well with the numerical result in Fig. \ref{fig:sqp}(a).

Next we consider the current of the LTQD at the AQP: $(011)$, $(012)$, $(021)$, and $(111)$.  We again use a five-level Hubbard Hamiltonian for transport calculation.  We put  $S$ orbital in the left dot, and  $S$ and $P$ orbitals in the central dot and the right dot.
In weak tunnel coupling limit, the 4 charge configurations should be on resonance, and we set $E_{1} = -V-V'$ and $E_{2} = E_{3} = -U-V$.  
Figure \ref{fig:aqp}(a) shows the $I (V_{sd})$ of the TQD near the AQP.  As expected, the current of the TQD near the AQP is very different under the two bias directions.  Figure \ref{fig:aqp}(b) confirms the spin blockade where the current is severely suppressed in the positive bias direction in Fig. \ref{fig:aqp}(a).  This phenomenon is in close analogy to the case of a DQD, and can be easily explained.  In the positive bias, electron is  injected from the left dot. The transition from $(0,1,1)$ triplet to a $(1,1,1)$ spin 3/2 state does not require the formation of on-site triplet. In the negative bias direction, electron is injected from the right and transition from $(0,1,1)$ triplet state to a spin 3/2 state requires the formation of an on-site triplet in the right dot. Therefore, before the bias threshold $\mu_R^* = E(0,1,2^*) - E(0,1,1)$, no spin blockade is expected to be formed. $(0,1,2^*)$ represents  charge configuration in which one of the electrons occupies the $P$ orbital in dot $3$.  When the applied bias exceeds the threshold, the condition $E(0,1,2^*) \geq E(0,2^*,1) \geq E(1,1,1)$ is also satisfied.  Thus, either due to resonant tunneling or inelastic process, this additional electron can always be removed from the right lead.  Therefore, there is no spin blockade in the negative bias direction.  We remark that the spin blockade at AQP can be formed without the assistance of any relaxation mechanism.    So the spin blockade of a TQD at AQP is almost identical to the spin blockade in a DQD.  Figure \ref{fig:aqp2} presents the transport triangle in the parameter space $(E_1,E_2=E_3)$ at a positive bias.  In this figure, the light trail at the tip of the triangle is proportional to the on-site singlet-triplet gap in the central dot.  This transport triangle, although generated under the specific condition $E_2=E_3$, provides similar information that one can extract from the transport triangle for the DQD.

\section{Conclusion} \label{sec:CON}
We presented a theory of electronic properties and transport through a LTQD around QPs. We showed that the spin blockade could serve as a spectroscopic tool for the detection of different spin states. Two different QPs containing the $(1,1,1)$ configuration were discussed. A multi-band Hubbard model with five levels was used to describe the electronic properties and investigate the spin blockade phenomenon in the LTQD. For both QPs, strong current suppression and negative differential conductance were predicted. At the SQP, suppression in conductance was obtained under two different source-drain bias regimes.  When the bias is small, the electronic transport involving spin-3/2 states takes place either via the $S$ or $P$ orbitals in the edge dot with comparable amplitude and results in a destructive interference. In high bias regime where electron tunnels onto the $P$ orbital in the edge dot, spin blockade is facilitated by spin-conserving relaxation mechanisms, such as interaction with LA phonons studied here, and formation of the trap state. At the SQP, the spin blockade phenomenon is bi-directional, in contrast with the spin blockade in a DQD. We also  discussed spin blockade at the AQP. The spin blockade formation and lifting in this case is in close analogy to the DQD case. The formation of the spin blockade does not involve any on-site triplets, only the lifting of the spin blockade requires the access to the on-site triplet states in the transport window.  Similar to the  DQD, the spin blockade phenomenon at AQP only occurs only in one of the bias directions.

\section{Acknowledgment} \label{sec:Ack}
The authors thank L.Gaudreau, G. Granger, A. Sachrajda for discussion and NSERC, QUANTUMWORKS, CIFAR, NRC-CNRS CRP, NRC-NSERC-BDC Nanotechnology project and OGS for financial support.


\begin{thebibliography}{40}
\expandafter\ifx\csname natexlab\endcsname\relax\def\natexlab#1{#1}\fi
\expandafter\ifx\csname bibnamefont\endcsname\relax
  \def\bibnamefont#1{#1}\fi
\expandafter\ifx\csname bibfnamefont\endcsname\relax
  \def\bibfnamefont#1{#1}\fi
\expandafter\ifx\csname citenamefont\endcsname\relax
  \def\citenamefont#1{#1}\fi
\expandafter\ifx\csname url\endcsname\relax
  \def\url#1{\texttt{#1}}\fi
\expandafter\ifx\csname urlprefix\endcsname\relax\def\urlprefix{URL }\fi
\providecommand{\bibinfo}[2]{#2}
\providecommand{\eprint}[2][]{\url{#2}}

\bibitem[{\citenamefont{Ciorga et~al.}(2000)\citenamefont{Ciorga, Sachrajda,
  Hawrylak, Gould, Zawadzki, Jullian, Feng, and
  Wasilewski}}]{ciorga_sachrajda_prb2000}
\bibinfo{author}{\bibfnamefont{M.}~\bibnamefont{Ciorga}},
  \bibinfo{author}{\bibfnamefont{A.~S.} \bibnamefont{Sachrajda}},
  \bibinfo{author}{\bibfnamefont{P.}~\bibnamefont{Hawrylak}},
  \bibinfo{author}{\bibfnamefont{C.}~\bibnamefont{Gould}},
  \bibinfo{author}{\bibfnamefont{P.}~\bibnamefont{Zawadzki}},
  \bibinfo{author}{\bibfnamefont{S.}~\bibnamefont{Jullian}},
  \bibinfo{author}{\bibfnamefont{Y.}~\bibnamefont{Feng}}, \bibnamefont{and}
  \bibinfo{author}{\bibfnamefont{Z.}~\bibnamefont{Wasilewski}},
  \bibinfo{journal}{Phys. Rev. B} \textbf{\bibinfo{volume}{61}},
  \bibinfo{pages}{R16315} (\bibinfo{year}{2000}).

\bibitem[{\citenamefont{Hawrylak}(1999)}]{hawrylak_prb1999}
\bibinfo{author}{\bibfnamefont{P.}~\bibnamefont{Hawrylak}},
  \bibinfo{journal}{Phys. Rev. B} \textbf{\bibinfo{volume}{60}},
  \bibinfo{pages}{5597} (\bibinfo{year}{1999}).

\bibitem[{\citenamefont{Hanson et~al.}(2007)\citenamefont{Hanson, Petta,
  Tarucha, and Vandersypen}}]{hanson_petta_rmp2007}
\bibinfo{author}{\bibfnamefont{R.}~\bibnamefont{Hanson}},
  \bibinfo{author}{\bibfnamefont{J.~R.} \bibnamefont{Petta}},
  \bibinfo{author}{\bibfnamefont{S.}~\bibnamefont{Tarucha}}, \bibnamefont{and}
  \bibinfo{author}{\bibfnamefont{L.~M.~K.} \bibnamefont{Vandersypen}},
  \bibinfo{journal}{Rev. Mod. Phys.} \textbf{\bibinfo{volume}{79}},
  \bibinfo{pages}{1217} (\bibinfo{year}{2007}).

\bibitem[{\citenamefont{Koppens et~al.}(2006)\citenamefont{Koppens, Buizert,
  Tielrooij, and Vink}}]{koppens_buizert_nat2006}
\bibinfo{author}{\bibfnamefont{F.}~\bibnamefont{Koppens}},
  \bibinfo{author}{\bibfnamefont{C.}~\bibnamefont{Buizert}},
  \bibinfo{author}{\bibfnamefont{K.}~\bibnamefont{Tielrooij}},
  \bibnamefont{and} \bibinfo{author}{\bibfnamefont{I.}~\bibnamefont{Vink}},
  \bibinfo{journal}{Nature} \textbf{\bibinfo{volume}{442}},
  \bibinfo{pages}{766} (\bibinfo{year}{2006}).

\bibitem[{\citenamefont{Nowack et~al.}(2007)\citenamefont{Nowack, Koppens,
  Nazarov, and Vandersypen}}]{nowack_koppens_sci2007}
\bibinfo{author}{\bibfnamefont{K.~C.} \bibnamefont{Nowack}},
  \bibinfo{author}{\bibfnamefont{F.~H.~L.} \bibnamefont{Koppens}},
  \bibinfo{author}{\bibfnamefont{Y.~V.} \bibnamefont{Nazarov}},
  \bibnamefont{and} \bibinfo{author}{\bibfnamefont{L.~M.~K.}
  \bibnamefont{Vandersypen}}, \bibinfo{journal}{Science}
  \textbf{\bibinfo{volume}{318}}, \bibinfo{pages}{1430} (\bibinfo{year}{2007}).

\bibitem[{\citenamefont{Pioro-Ladri{\`e}re
  et~al.}(2008)\citenamefont{Pioro-Ladri{\`e}re, Obata, Tokura, Shin, Kubo,
  Yoshida, Taniyama, and Tarucha}}]{pioroLadriere_obata_natphys2008}
\bibinfo{author}{\bibfnamefont{M.}~\bibnamefont{Pioro-Ladri{\`e}re}},
  \bibinfo{author}{\bibfnamefont{T.}~\bibnamefont{Obata}},
  \bibinfo{author}{\bibfnamefont{Y.}~\bibnamefont{Tokura}},
  \bibinfo{author}{\bibfnamefont{Y.~S.} \bibnamefont{Shin}},
  \bibinfo{author}{\bibfnamefont{T.}~\bibnamefont{Kubo}},
  \bibinfo{author}{\bibfnamefont{K.}~\bibnamefont{Yoshida}},
  \bibinfo{author}{\bibfnamefont{T.}~\bibnamefont{Taniyama}}, \bibnamefont{and}
  \bibinfo{author}{\bibfnamefont{S.}~\bibnamefont{Tarucha}},
  \bibinfo{journal}{Nat Phys} \textbf{\bibinfo{volume}{4}},
  \bibinfo{pages}{776} (\bibinfo{year}{2008}).

\bibitem[{\citenamefont{Gaudreau et~al.}(2009)\citenamefont{Gaudreau, Kam,
  Granger, Studenikin, Zawadzki, and Sachrajda}}]{gaudreau_kam_apl2009}
\bibinfo{author}{\bibfnamefont{L.}~\bibnamefont{Gaudreau}},
  \bibinfo{author}{\bibfnamefont{A.}~\bibnamefont{Kam}},
  \bibinfo{author}{\bibfnamefont{G.}~\bibnamefont{Granger}},
  \bibinfo{author}{\bibfnamefont{S.~A.} \bibnamefont{Studenikin}},
  \bibinfo{author}{\bibfnamefont{P.}~\bibnamefont{Zawadzki}}, \bibnamefont{and}
  \bibinfo{author}{\bibfnamefont{A.~S.} \bibnamefont{Sachrajda}},
  \bibinfo{journal}{Appl. Phys. Lett.} \textbf{\bibinfo{volume}{95}},
  \bibinfo{pages}{193101} (\bibinfo{year}{2009}).

\bibitem[{\citenamefont{Laird et~al.}(2010)\citenamefont{Laird, Taylor,
  DiVincenzo, Marcus, Hanson, and Gossard}}]{laird_taylor_prb2010}
\bibinfo{author}{\bibfnamefont{E.~A.} \bibnamefont{Laird}},
  \bibinfo{author}{\bibfnamefont{J.~M.} \bibnamefont{Taylor}},
  \bibinfo{author}{\bibfnamefont{D.~P.} \bibnamefont{DiVincenzo}},
  \bibinfo{author}{\bibfnamefont{C.~M.} \bibnamefont{Marcus}},
  \bibinfo{author}{\bibfnamefont{M.~P.} \bibnamefont{Hanson}},
  \bibnamefont{and} \bibinfo{author}{\bibfnamefont{A.~C.}
  \bibnamefont{Gossard}}, \bibinfo{journal}{Phys. Rev. B}
  \textbf{\bibinfo{volume}{82}}, \bibinfo{pages}{075403}
  (\bibinfo{year}{2010}).

\bibitem[{\citenamefont{Korkusinski and
  Hawrylak}(2008)}]{korkusinski_hawrylak_wsr2007}
\bibinfo{author}{\bibfnamefont{M.}~\bibnamefont{Korkusinski}} \bibnamefont{and}
  \bibinfo{author}{\bibfnamefont{P.}~\bibnamefont{Hawrylak}},
  \bibinfo{journal}{in ''Semiconductor quantum bits'', ed. by O. Benson and F.
  Henneberger, World Scientific}  (\bibinfo{year}{2008}).

\bibitem[{\citenamefont{Petta et~al.}(2005)\citenamefont{Petta, Johnson,
  Taylor, Laird, Yacoby, Lukin, Marcus, and Hanson}}]{petta_johnson_sci2005}
\bibinfo{author}{\bibfnamefont{J.}~\bibnamefont{Petta}},
  \bibinfo{author}{\bibfnamefont{A.}~\bibnamefont{Johnson}},
  \bibinfo{author}{\bibfnamefont{J.}~\bibnamefont{Taylor}},
  \bibinfo{author}{\bibfnamefont{E.}~\bibnamefont{Laird}},
  \bibinfo{author}{\bibfnamefont{A.}~\bibnamefont{Yacoby}},
  \bibinfo{author}{\bibfnamefont{M.~D.} \bibnamefont{Lukin}},
  \bibinfo{author}{\bibfnamefont{C.~M.} \bibnamefont{Marcus}},
  \bibnamefont{and} \bibinfo{author}{\bibfnamefont{M.~P.}
  \bibnamefont{Hanson}}, \bibinfo{journal}{Science}
  \textbf{\bibinfo{volume}{309}}, \bibinfo{pages}{2180} (\bibinfo{year}{2005}).

\bibitem[{\citenamefont{Johnson et~al.}(2005)\citenamefont{Johnson, Petta,
  Marcus, Hanson, and Gossard}}]{johnson_petta_prb2005}
\bibinfo{author}{\bibfnamefont{A.~C.} \bibnamefont{Johnson}},
  \bibinfo{author}{\bibfnamefont{J.~R.} \bibnamefont{Petta}},
  \bibinfo{author}{\bibfnamefont{C.~M.} \bibnamefont{Marcus}},
  \bibinfo{author}{\bibfnamefont{M.~P.} \bibnamefont{Hanson}},
  \bibnamefont{and} \bibinfo{author}{\bibfnamefont{A.~C.}
  \bibnamefont{Gossard}}, \bibinfo{journal}{Phys. Rev. B}
  \textbf{\bibinfo{volume}{72}}, \bibinfo{pages}{165308}
  (\bibinfo{year}{2005}).

\bibitem[{\citenamefont{Ono et~al.}(2002)\citenamefont{Ono, Austing, Tokura,
  and Tarucha}}]{ono_austing_sci2002}
\bibinfo{author}{\bibfnamefont{K.}~\bibnamefont{Ono}},
  \bibinfo{author}{\bibfnamefont{D.~G.} \bibnamefont{Austing}},
  \bibinfo{author}{\bibfnamefont{Y.}~\bibnamefont{Tokura}}, \bibnamefont{and}
  \bibinfo{author}{\bibfnamefont{S.}~\bibnamefont{Tarucha}},
  \bibinfo{journal}{Science} \textbf{\bibinfo{volume}{297}},
  \bibinfo{pages}{1313} (\bibinfo{year}{2002}).

\bibitem[{\citenamefont{Granger et~al.}(2010)\citenamefont{Granger, Gaudreau,
  Kam, Pioro-Ladri{\`e}re, Studenikin, Wasilewski, Zawadzki, and
  Sachrajda}}]{granger_gaudreau_prb2010}
\bibinfo{author}{\bibfnamefont{G.}~\bibnamefont{Granger}},
  \bibinfo{author}{\bibfnamefont{L.}~\bibnamefont{Gaudreau}},
  \bibinfo{author}{\bibfnamefont{A.}~\bibnamefont{Kam}},
  \bibinfo{author}{\bibfnamefont{M.}~\bibnamefont{Pioro-Ladri{\`e}re}},
  \bibinfo{author}{\bibfnamefont{S.~A.} \bibnamefont{Studenikin}},
  \bibinfo{author}{\bibfnamefont{Z.~R.} \bibnamefont{Wasilewski}},
  \bibinfo{author}{\bibfnamefont{P.}~\bibnamefont{Zawadzki}}, \bibnamefont{and}
  \bibinfo{author}{\bibfnamefont{A.~S.} \bibnamefont{Sachrajda}},
  \bibinfo{journal}{Phys. Rev. B} \textbf{\bibinfo{volume}{82}},
  \bibinfo{pages}{075304} (\bibinfo{year}{2010}).

\bibitem[{\citenamefont{Korkusinski et~al.}(2007)\citenamefont{Korkusinski,
  Gimenez, Hawrylak, Gaudreau, Studenikin, and
  Sachrajda}}]{korkusinski_gimenez_prb2007}
\bibinfo{author}{\bibfnamefont{M.}~\bibnamefont{Korkusinski}},
  \bibinfo{author}{\bibfnamefont{I.~P.} \bibnamefont{Gimenez}},
  \bibinfo{author}{\bibfnamefont{P.}~\bibnamefont{Hawrylak}},
  \bibinfo{author}{\bibfnamefont{L.}~\bibnamefont{Gaudreau}},
  \bibinfo{author}{\bibfnamefont{S.~A.} \bibnamefont{Studenikin}},
  \bibnamefont{and} \bibinfo{author}{\bibfnamefont{A.~S.}
  \bibnamefont{Sachrajda}}, \bibinfo{journal}{Phys. Rev. B}
  \textbf{\bibinfo{volume}{75}}, \bibinfo{pages}{115301}
  (\bibinfo{year}{2007}).

\bibitem[{\citenamefont{Grove-Rasmussen
  et~al.}(2008)\citenamefont{Grove-Rasmussen, J{\o}rgensen, Hayashi, Lindelof,
  and Fujisawa}}]{groveRasmussen_jorgensen_nanolett2008}
\bibinfo{author}{\bibfnamefont{K.}~\bibnamefont{Grove-Rasmussen}},
  \bibinfo{author}{\bibfnamefont{H.~I.} \bibnamefont{J{\o}rgensen}},
  \bibinfo{author}{\bibfnamefont{T.}~\bibnamefont{Hayashi}},
  \bibinfo{author}{\bibfnamefont{P.~E.} \bibnamefont{Lindelof}},
  \bibnamefont{and} \bibinfo{author}{\bibfnamefont{T.}~\bibnamefont{Fujisawa}},
  \bibinfo{journal}{Nano Lett.} \textbf{\bibinfo{volume}{8}},
  \bibinfo{pages}{1055} (\bibinfo{year}{2008}).

\bibitem[{\citenamefont{Gaudreau et~al.}(2011)\citenamefont{Gaudreau, Granger,
  Kam, Aers, Studenikin, Zawadzki, Pioro-Ladri\'{e}re, Wasilewski, and
  Sachrajda}}]{gaudreau_granger_arxiv2011}
\bibinfo{author}{\bibfnamefont{L.}~\bibnamefont{Gaudreau}},
  \bibinfo{author}{\bibfnamefont{G.}~\bibnamefont{Granger}},
  \bibinfo{author}{\bibfnamefont{A.}~\bibnamefont{Kam}},
  \bibinfo{author}{\bibfnamefont{G.~C.} \bibnamefont{Aers}},
  \bibinfo{author}{\bibfnamefont{S.~A.} \bibnamefont{Studenikin}},
  \bibinfo{author}{\bibfnamefont{P.}~\bibnamefont{Zawadzki}},
  \bibinfo{author}{\bibfnamefont{M.}~\bibnamefont{Pioro-Ladri\'{e}re}},
  \bibinfo{author}{\bibfnamefont{Z.~R.} \bibnamefont{Wasilewski}},
  \bibnamefont{and} \bibinfo{author}{\bibfnamefont{A.~S.}
  \bibnamefont{Sachrajda}}, \bibinfo{journal}{arxiv:1106.3518v1}
  (\bibinfo{year}{2011}).

\bibitem[{\citenamefont{Shim et~al.}(2009)\citenamefont{Shim, Delgado, and
  Hawrylak}}]{shim_delgado_prb2009}
\bibinfo{author}{\bibfnamefont{Y.~P.} \bibnamefont{Shim}},
  \bibinfo{author}{\bibfnamefont{F.}~\bibnamefont{Delgado}}, \bibnamefont{and}
  \bibinfo{author}{\bibfnamefont{P.}~\bibnamefont{Hawrylak}},
  \bibinfo{journal}{Phys. Rev. B} \textbf{\bibinfo{volume}{80}},
  \bibinfo{pages}{115305} (\bibinfo{year}{2009}).

\bibitem[{\citenamefont{Delgado et~al.}(2008)\citenamefont{Delgado, Shim,
  Korkusinski, Gaudreau, Studenikin, Sachrajda, and P.}}]{delgado_shim_prl2008}
\bibinfo{author}{\bibfnamefont{F.}~\bibnamefont{Delgado}},
  \bibinfo{author}{\bibfnamefont{Y.~P.} \bibnamefont{Shim}},
  \bibinfo{author}{\bibfnamefont{M.}~\bibnamefont{Korkusinski}},
  \bibinfo{author}{\bibfnamefont{L.}~\bibnamefont{Gaudreau}},
  \bibinfo{author}{\bibfnamefont{S.~A.} \bibnamefont{Studenikin}},
  \bibinfo{author}{\bibnamefont{Sachrajda}}, \bibnamefont{and}
  \bibinfo{author}{\bibfnamefont{H.}~\bibnamefont{P.}}, \bibinfo{journal}{Phys.
  Rev. Lett.} \textbf{\bibinfo{volume}{101}}, \bibinfo{pages}{226810}
  (\bibinfo{year}{2008}).

\bibitem[{\citenamefont{DiVincenzo et~al.}(2000)\citenamefont{DiVincenzo,
  Bacon, Kempe, and Burkard}}]{diVincenzo_bacon_nat2000}
\bibinfo{author}{\bibfnamefont{D.}~\bibnamefont{DiVincenzo}},
  \bibinfo{author}{\bibfnamefont{D.}~\bibnamefont{Bacon}},
  \bibinfo{author}{\bibfnamefont{J.}~\bibnamefont{Kempe}}, \bibnamefont{and}
  \bibinfo{author}{\bibfnamefont{G.}~\bibnamefont{Burkard}},
  \bibinfo{journal}{Nature} \textbf{\bibinfo{volume}{408}},
  \bibinfo{pages}{339} (\bibinfo{year}{2000}).

\bibitem[{\citenamefont{Hawrylak and
  Korkusinski}(2005)}]{hawrylak_korkusinski_ssc2005}
\bibinfo{author}{\bibfnamefont{P.}~\bibnamefont{Hawrylak}} \bibnamefont{and}
  \bibinfo{author}{\bibfnamefont{M.}~\bibnamefont{Korkusinski}},
  \bibinfo{journal}{Solid State Communications} \textbf{\bibinfo{volume}{136}},
  \bibinfo{pages}{508} (\bibinfo{year}{2005}).

\bibitem[{\citenamefont{Hsieh and Hawrylak}(2010)}]{hsieh_hawrylak_prb2010}
\bibinfo{author}{\bibfnamefont{C.-Y.} \bibnamefont{Hsieh}} \bibnamefont{and}
  \bibinfo{author}{\bibfnamefont{P.}~\bibnamefont{Hawrylak}},
  \bibinfo{journal}{Phys. Rev. B} \textbf{\bibinfo{volume}{82}},
  \bibinfo{pages}{205311} (\bibinfo{year}{2010}).

\bibitem[{\citenamefont{Shim et~al.}(2010)\citenamefont{Shim, Sharma, Hsieh,
  and Hawrylak}}]{shim_sharma_ssc2010}
\bibinfo{author}{\bibfnamefont{Y.-P.} \bibnamefont{Shim}},
  \bibinfo{author}{\bibfnamefont{A.}~\bibnamefont{Sharma}},
  \bibinfo{author}{\bibfnamefont{C.-Y.} \bibnamefont{Hsieh}}, \bibnamefont{and}
  \bibinfo{author}{\bibfnamefont{P.}~\bibnamefont{Hawrylak}},
  \bibinfo{journal}{Solid State Commun.} \textbf{\bibinfo{volume}{150}},
  \bibinfo{pages}{2065} (\bibinfo{year}{2010}).

\bibitem[{\citenamefont{Shim and Hawrylak}(2008)}]{shim_hawrylak_prb2008}
\bibinfo{author}{\bibfnamefont{Y.-P.} \bibnamefont{Shim}} \bibnamefont{and}
  \bibinfo{author}{\bibfnamefont{P.}~\bibnamefont{Hawrylak}},
  \bibinfo{journal}{Phys. Rev. B} \textbf{\bibinfo{volume}{78}},
  \bibinfo{pages}{165317} (\bibinfo{year}{2008}).

\bibitem[{\citenamefont{R{\"o}thlisberger
  et~al.}(2008)\citenamefont{R{\"o}thlisberger, Lehmann, Saraga, Traber, and
  Loss}}]{rothlisberger_lehmann_prl2008}
\bibinfo{author}{\bibfnamefont{B.}~\bibnamefont{R{\"o}thlisberger}},
  \bibinfo{author}{\bibfnamefont{J.}~\bibnamefont{Lehmann}},
  \bibinfo{author}{\bibfnamefont{D.~S.} \bibnamefont{Saraga}},
  \bibinfo{author}{\bibfnamefont{P.}~\bibnamefont{Traber}}, \bibnamefont{and}
  \bibinfo{author}{\bibfnamefont{D.}~\bibnamefont{Loss}},
  \bibinfo{journal}{Phys. Rev. Lett.} \textbf{\bibinfo{volume}{100}},
  \bibinfo{pages}{100502} (\bibinfo{year}{2008}).

\bibitem[{\citenamefont{Sharma and Hawrylak}(2011)}]{sharma_hawrylak_prb2011}
\bibinfo{author}{\bibfnamefont{A.}~\bibnamefont{Sharma}} \bibnamefont{and}
  \bibinfo{author}{\bibfnamefont{P.}~\bibnamefont{Hawrylak}},
  \bibinfo{journal}{Phys. Rev. B} \textbf{\bibinfo{volume}{83}},
  \bibinfo{pages}{125311} (\bibinfo{year}{2011}).

\bibitem[{\citenamefont{Ingersent et~al.}(2005)\citenamefont{Ingersent, Ludwig,
  and Affleck}}]{ingersent_ludwig_prl2005}
\bibinfo{author}{\bibfnamefont{K.}~\bibnamefont{Ingersent}},
  \bibinfo{author}{\bibfnamefont{A.~W.~W.} \bibnamefont{Ludwig}},
  \bibnamefont{and} \bibinfo{author}{\bibfnamefont{I.}~\bibnamefont{Affleck}},
  \bibinfo{journal}{Phys. Rev. Lett.} \textbf{\bibinfo{volume}{95}},
  \bibinfo{pages}{257204} (\bibinfo{year}{2005}).

\bibitem[{\citenamefont{{\v Z}itko and Bon{\v c}a}(2007)}]{zitko_bonca_prl2007}
\bibinfo{author}{\bibfnamefont{R.}~\bibnamefont{{\v Z}itko}} \bibnamefont{and}
  \bibinfo{author}{\bibfnamefont{J.}~\bibnamefont{Bon{\v c}a}},
  \bibinfo{journal}{Phys. Rev. Lett.} \textbf{\bibinfo{volume}{98}},
  \bibinfo{pages}{047203} (\bibinfo{year}{2007}).

\bibitem[{\citenamefont{Lobos and Aligia}(2006)}]{lobos_aligia_prb2006}
\bibinfo{author}{\bibfnamefont{A.~M.} \bibnamefont{Lobos}} \bibnamefont{and}
  \bibinfo{author}{\bibfnamefont{A.~A.} \bibnamefont{Aligia}},
  \bibinfo{journal}{Phys. Rev. B} \textbf{\bibinfo{volume}{74}},
  \bibinfo{pages}{165417} (\bibinfo{year}{2006}).

\bibitem[{\citenamefont{Greentree
  et~al.}(2004{\natexlab{a}})\citenamefont{Greentree, Cole, Hamilton, and
  Hollenberg}}]{greentree_cole_prb2004}
\bibinfo{author}{\bibfnamefont{A.~D.} \bibnamefont{Greentree}},
  \bibinfo{author}{\bibfnamefont{J.~H.} \bibnamefont{Cole}},
  \bibinfo{author}{\bibfnamefont{A.~R.} \bibnamefont{Hamilton}},
  \bibnamefont{and} \bibinfo{author}{\bibfnamefont{L.~C.~L.}
  \bibnamefont{Hollenberg}}, \bibinfo{journal}{Phys. Rev. B}
  \textbf{\bibinfo{volume}{70}}, \bibinfo{pages}{235317}
  (\bibinfo{year}{2004}{\natexlab{a}}).

\bibitem[{\citenamefont{Emary}(2007)}]{emary_prb2007}
\bibinfo{author}{\bibfnamefont{C.}~\bibnamefont{Emary}},
  \bibinfo{journal}{Phys. Rev. B} \textbf{\bibinfo{volume}{76}},
  \bibinfo{pages}{245319} (\bibinfo{year}{2007}).

\bibitem[{\citenamefont{PuertoGimenez et~al.}(2007)\citenamefont{PuertoGimenez,
  Korkusinski, and Hawrylak}}]{gimenez_korkusinski_prb2007}
\bibinfo{author}{\bibfnamefont{I.}~\bibnamefont{PuertoGimenez}},
  \bibinfo{author}{\bibfnamefont{M.}~\bibnamefont{Korkusinski}},
  \bibnamefont{and} \bibinfo{author}{\bibfnamefont{P.}~\bibnamefont{Hawrylak}},
  \bibinfo{journal}{Phys. Rev. B} \textbf{\bibinfo{volume}{76}},
  \bibinfo{pages}{075336} (\bibinfo{year}{2007}).

\bibitem[{\citenamefont{Gimenez et~al.}(2009)\citenamefont{Gimenez, Hsieh,
  Korkusinski, and Hawrylak}}]{gimenez_hsieh_prb2009}
\bibinfo{author}{\bibfnamefont{I.~P.} \bibnamefont{Gimenez}},
  \bibinfo{author}{\bibfnamefont{C.-Y.} \bibnamefont{Hsieh}},
  \bibinfo{author}{\bibfnamefont{M.}~\bibnamefont{Korkusinski}},
  \bibnamefont{and} \bibinfo{author}{\bibfnamefont{P.}~\bibnamefont{Hawrylak}},
  \bibinfo{journal}{Phys. Rev. B} \textbf{\bibinfo{volume}{79}},
  \bibinfo{pages}{205311} (\bibinfo{year}{2009}).

\bibitem[{\citenamefont{Muralidharan and
  Datta}(2007)}]{muralidharan_datta_prb2007}
\bibinfo{author}{\bibfnamefont{B.}~\bibnamefont{Muralidharan}}
  \bibnamefont{and} \bibinfo{author}{\bibfnamefont{S.}~\bibnamefont{Datta}},
  \bibinfo{journal}{Phys. Rev. B} \textbf{\bibinfo{volume}{76}},
  \bibinfo{pages}{035432} (\bibinfo{year}{2007}).

\bibitem[{\citenamefont{Florescu and
  Hawrylak}(2006)}]{florescu_hawrylak_prb2006}
\bibinfo{author}{\bibfnamefont{M.}~\bibnamefont{Florescu}} \bibnamefont{and}
  \bibinfo{author}{\bibfnamefont{P.}~\bibnamefont{Hawrylak}},
  \bibinfo{journal}{Phys. Rev. B} \textbf{\bibinfo{volume}{73}},
  \bibinfo{pages}{045304} (\bibinfo{year}{2006}).

\bibitem[{\citenamefont{I{\~n}arrea et~al.}(2007)\citenamefont{I{\~n}arrea,
  Platero, and MacDonald}}]{inarrea_platero_prb2008}
\bibinfo{author}{\bibfnamefont{J.}~\bibnamefont{I{\~n}arrea}},
  \bibinfo{author}{\bibfnamefont{G.}~\bibnamefont{Platero}}, \bibnamefont{and}
  \bibinfo{author}{\bibfnamefont{A.~H.} \bibnamefont{MacDonald}},
  \bibinfo{journal}{Phys. Rev. B} \textbf{\bibinfo{volume}{76}},
  \bibinfo{pages}{085329} (\bibinfo{year}{2007}).

\bibitem[{\citenamefont{Michaelis et~al.}(2006)\citenamefont{Michaelis, Emary,
  and Beenakker}}]{michaelis_emary_epl2006}
\bibinfo{author}{\bibfnamefont{B.}~\bibnamefont{Michaelis}},
  \bibinfo{author}{\bibfnamefont{C.}~\bibnamefont{Emary}}, \bibnamefont{and}
  \bibinfo{author}{\bibfnamefont{C.~W.~J.} \bibnamefont{Beenakker}},
  \bibinfo{journal}{Europhys. Lett.} \textbf{\bibinfo{volume}{73}},
  \bibinfo{pages}{677} (\bibinfo{year}{2006}).

\bibitem[{\citenamefont{Greentree
  et~al.}(2004{\natexlab{b}})\citenamefont{Greentree, Hamilton, and
  Green}}]{greentree_hamilton_prb2004}
\bibinfo{author}{\bibfnamefont{A.~D.} \bibnamefont{Greentree}},
  \bibinfo{author}{\bibfnamefont{A.~R.} \bibnamefont{Hamilton}},
  \bibnamefont{and} \bibinfo{author}{\bibfnamefont{F.}~\bibnamefont{Green}},
  \bibinfo{journal}{Phys. Rev. B} \textbf{\bibinfo{volume}{70}},
  \bibinfo{pages}{041305(R)} (\bibinfo{year}{2004}{\natexlab{b}}).

\bibitem[{\citenamefont{Vaz and Kyriakidis}(2008)}]{vaz_kyriakidis_jcp2009}
\bibinfo{author}{\bibfnamefont{E.}~\bibnamefont{Vaz}} \bibnamefont{and}
  \bibinfo{author}{\bibfnamefont{J.}~\bibnamefont{Kyriakidis}},
  \bibinfo{journal}{Journal of Chemical Physics}
  \textbf{\bibinfo{volume}{129}}, \bibinfo{pages}{024903}
  (\bibinfo{year}{2008}).

\bibitem[{\citenamefont{Hansen et~al.}(2008)\citenamefont{Hansen, Mujica, and
  Ratner}}]{hansen_mujica_nanolett2008}
\bibinfo{author}{\bibfnamefont{T.}~\bibnamefont{Hansen}},
  \bibinfo{author}{\bibfnamefont{V.}~\bibnamefont{Mujica}}, \bibnamefont{and}
  \bibinfo{author}{\bibfnamefont{M.~A.} \bibnamefont{Ratner}},
  \bibinfo{journal}{Nano Lett.} \textbf{\bibinfo{volume}{8}},
  \bibinfo{pages}{3525} (\bibinfo{year}{2008}).

\bibitem[{\citenamefont{Qassemi et~al.}(2009)\citenamefont{Qassemi, Coish, and
  Wilhelm}}]{qassemi_coish_prl2009}
\bibinfo{author}{\bibfnamefont{F.}~\bibnamefont{Qassemi}},
  \bibinfo{author}{\bibfnamefont{W.~A.} \bibnamefont{Coish}}, \bibnamefont{and}
  \bibinfo{author}{\bibfnamefont{F.~K.} \bibnamefont{Wilhelm}},
  \bibinfo{journal}{Phys. Rev. Lett.} \textbf{\bibinfo{volume}{102}},
  \bibinfo{pages}{176806} (\bibinfo{year}{2009}).

\end{thebibliography}



\newpage
\begin{figure}
\centering
\includegraphics[width=0.9\textwidth]{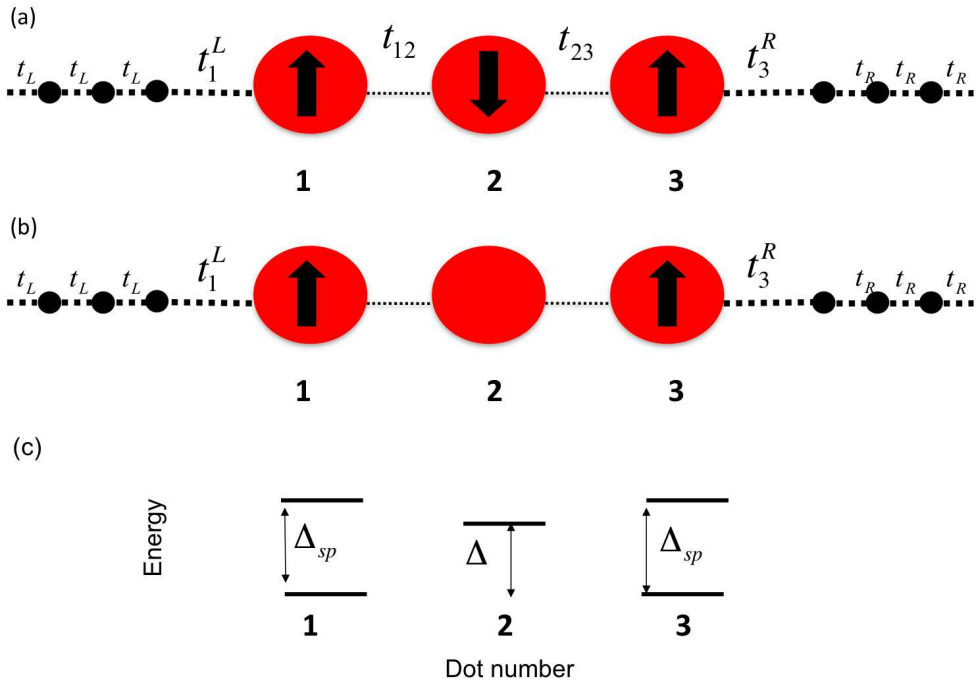}
\caption[Model of LTQD]{(Color Online) (a) Schematic picture of a LTQD with one electron spin each connected to leads. 
The leads are modelled with 1D tight binding chains.
(b) The TQD in $(1,0,1)$, two-electron configuration. (c) Schematic picture of single particle energy spectrum of a TQD when the central dot is biased with $\Delta$.  The gap $\Delta_{sp}$ denotes the energy difference between $S$ and $P$ orbitals on a dot.}
\label{fig:layout}
\end{figure}

\begin{figure}
\centering
\includegraphics[width=0.9\textwidth]{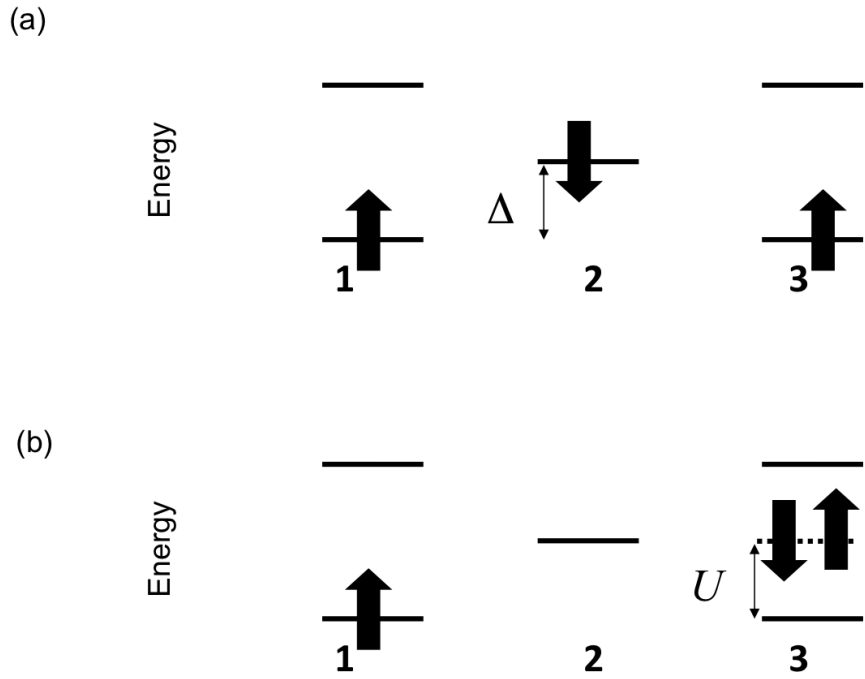}
\caption[Conditions of quadruple Point]{(Color Online) Two resonant three-electron configurations in a centrally biased TQD : (a) one of the (1,1,1) singly occupied configurations and  (b) one of the (1,0,2) doubly occupied configurations.  }
\label{fig:enres}
\end{figure}

\begin{figure}
\centering
\includegraphics[width=0.75\textwidth]{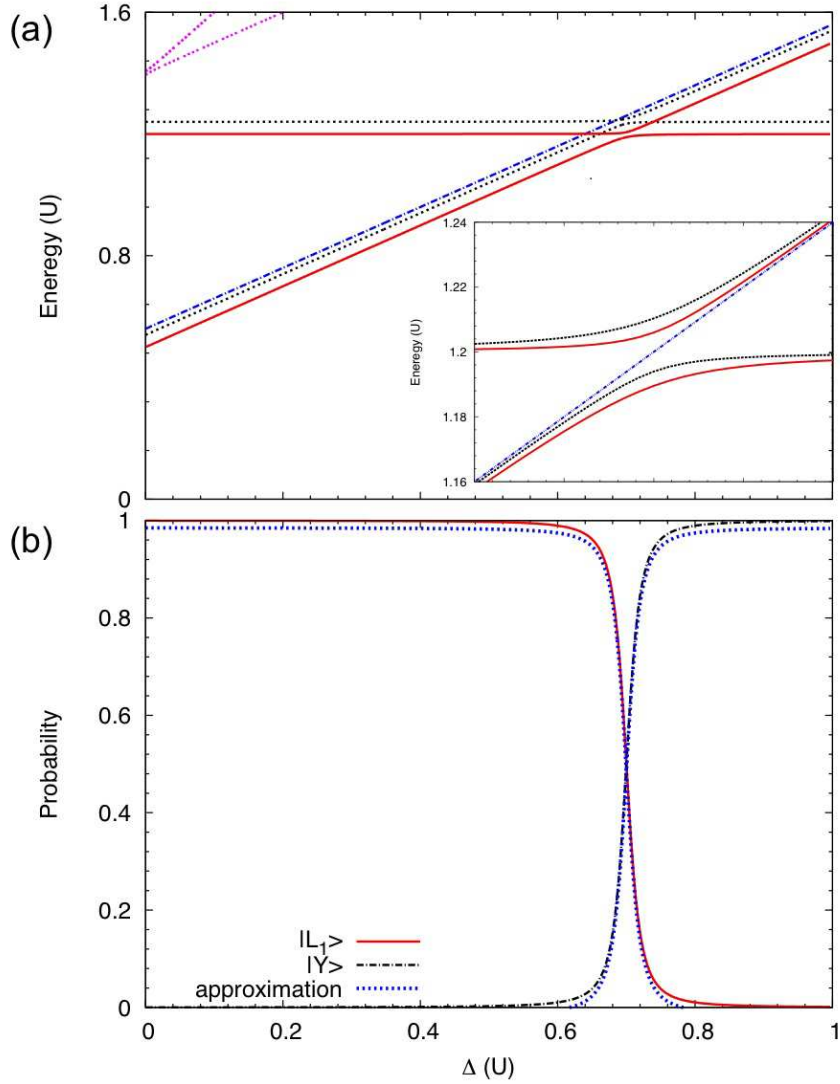}
\caption[Three-electron ground state analysis]{(Color Online) (a) The three electron molecule energy spectrum with total $S_y=1/2$
as a function of bias $\Delta$, obtained from the single-band Hubbard model.  As $\Delta$ increases, the five energy levels  anti-cross.  The blue curve, corresponding to total spin-3/2 state, does not interact with the other states. The energy levels are artificially shifted by a constant values for better visibility.  The inset shows the  energy levels near the anti-crossing point. (b) shows the projection of the ground state onto $\vert L_1 \rangle$ (the red curve) and $\vert Y \rangle$ (the black curve) states.  The wavefunction of the ground state is obtained from exact diagonalization of the single-band Hubbard model.  The blue-dashed lines provide the same information but obtained from the analytical approximation for $\vert L_1^+ \rangle = \cos(\theta) \ket{L_1} + \sin(\theta) \ket{Y} $ discussed in the text.}
\label{fig:sqp_cdoten}
\end{figure}

\begin{figure}
\centering
\includegraphics[width=0.9\textwidth]{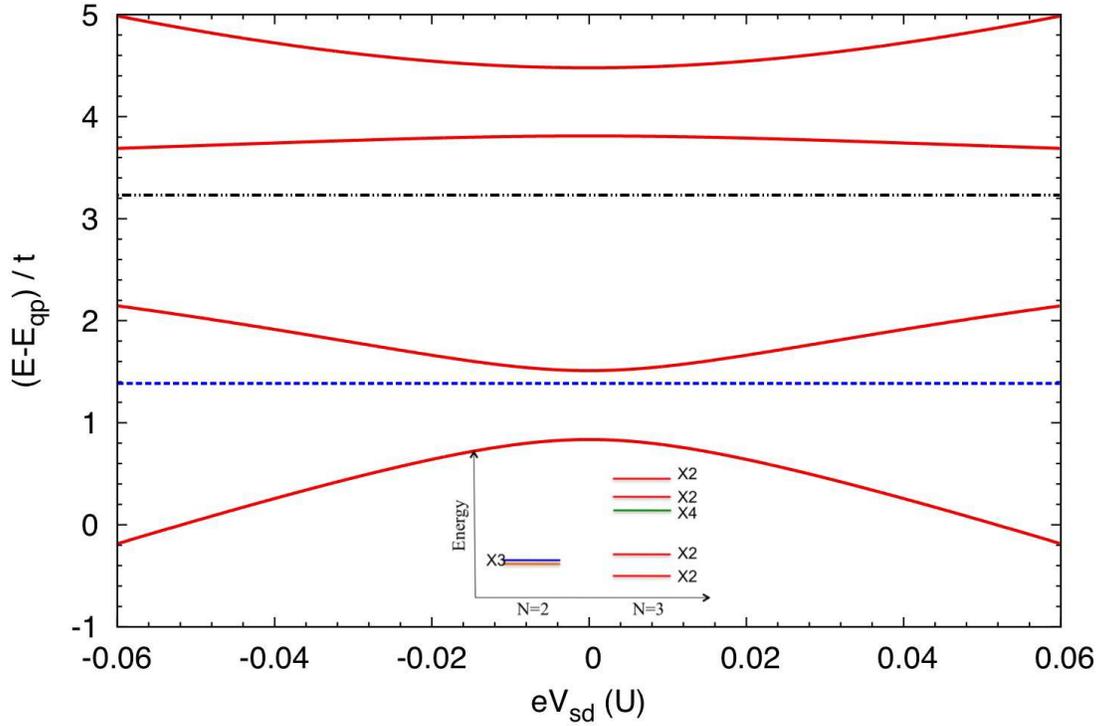}
\caption{(Color Online) The energy spectrum of a multi-band Hubbard model as a function of source-drain bias.  In the figure, the blue curve represents the four 2-electron states.  They are very close in energy and looks degenerate in the energy resolution present here.  The green curve is the quadruply degenerate spin-3/2 states.  The red curves are each doubly degenerate spin-1/2 states.  Inset: A summary of the states involved in the main figure at a particular value of $V_{sd}$.}
\label{fig:sqp_en}
\end{figure}

\begin{figure}
\centering
\includegraphics[width=0.8\textwidth, angle=270]{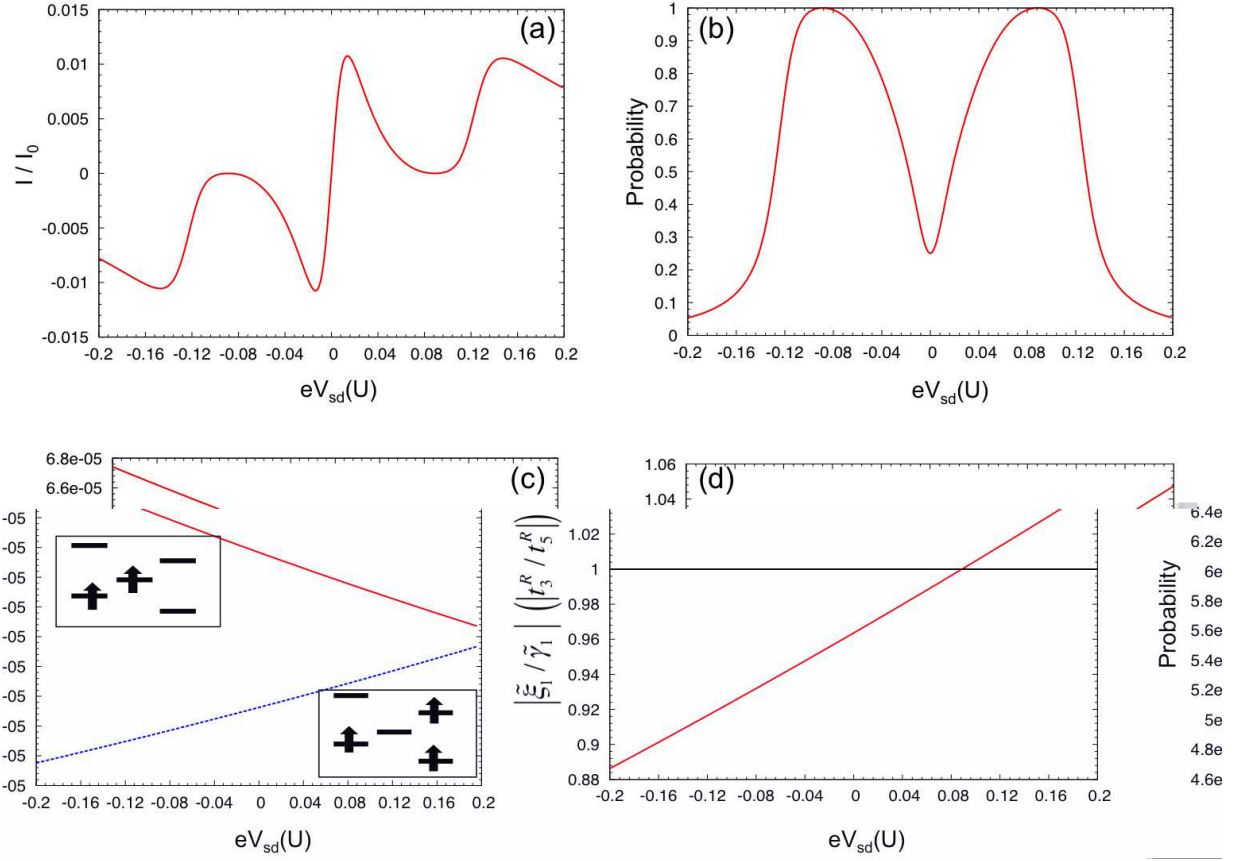}
\caption[$I-V$ curve at symmetric quadruple point]{(Color Online) (a) The current $I(V_{sd})$ as function of the applied source-drain bias $V_{sd}$ of a LTQD at the SQP in the low source-drain bias regime. Note zero current at $V_{sd}\approx \pm 0.08 $.  (b) The steady-state occupation probability of the spin-3/2 states as a function of $V_{sd}$.  Panels (a) and (b) together indicate that the spin-3/2 states are related to the bi-directional, quantum interference-based dark channel in a LTQD.  (c) Projection of the triplet state $\vert T^+ \rangle$ onto the configuration $\hat{c}^{\dag}_{1\uparrow}\hat{c}^{\dag}_{2\uparrow} \vert 0 \rangle$, and the projection of the spin-3/2 state $\vert 3/2 \rangle$ onto the configuration $\hat{c}^{\dag}_{1\uparrow}\hat{c}^{\dag}_{3\uparrow} \hat{c}^{\dag}_{5\uparrow} \vert 0 \rangle$.  (d) Ratio of matrix elements $\vert \frac{\eta_1} {\gamma_1} \vert$ (see text for the definition) in the unit of the ratio $\vert \frac{t^R_3}{t^R_5} \vert $.  Spin blockade is formed when the red curve  intercepts $y=1$ line in the figure. Panels (c) and (d) are presented to illustrate the formation of the spin blockade in the positive bias direction.   }
\label{fig:qmblockade}
\end{figure}

\begin{figure}
\centering
\includegraphics[width=0.7\textwidth]{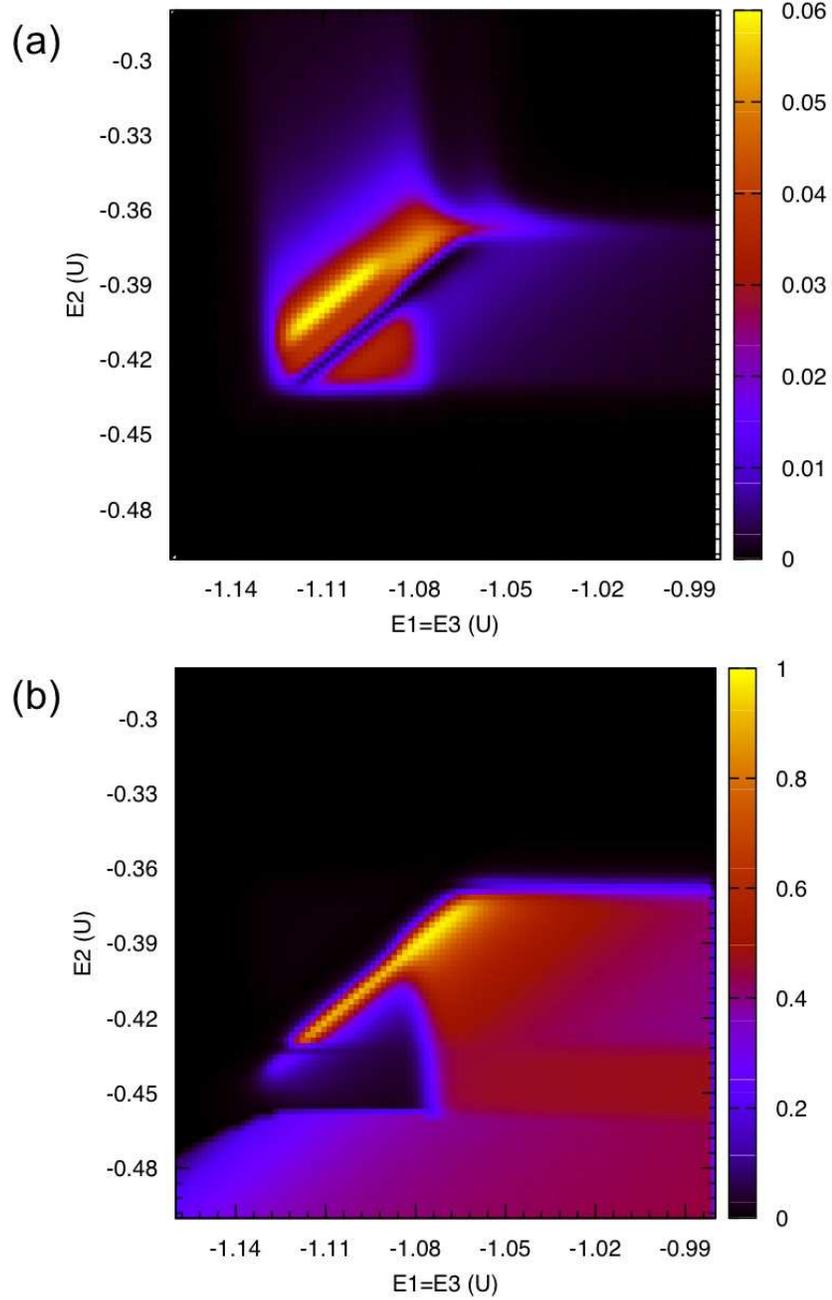}
\caption[Transport region in the on-site energy parameter space]{(Color Online) (a) Current of the LTQD in the parameter space of $(E_1=E_3,E_2)$.  The transport region manifests a rounded boundary, which indicate the states involves in the electronic transport are highly hybridized states.  The transport region is separated into two parts by a thin line of strong current suppression.  This is the region of dark channels. (b) The steady-state occupation probability for spin-3/2 states.  High amount of spin-3/2 states are found exactly where the currents vanishes in Panel (a).}
\label{fig:qmblockade2}
\end{figure}

\begin{figure}
\centering
\includegraphics[width=0.8\textwidth]{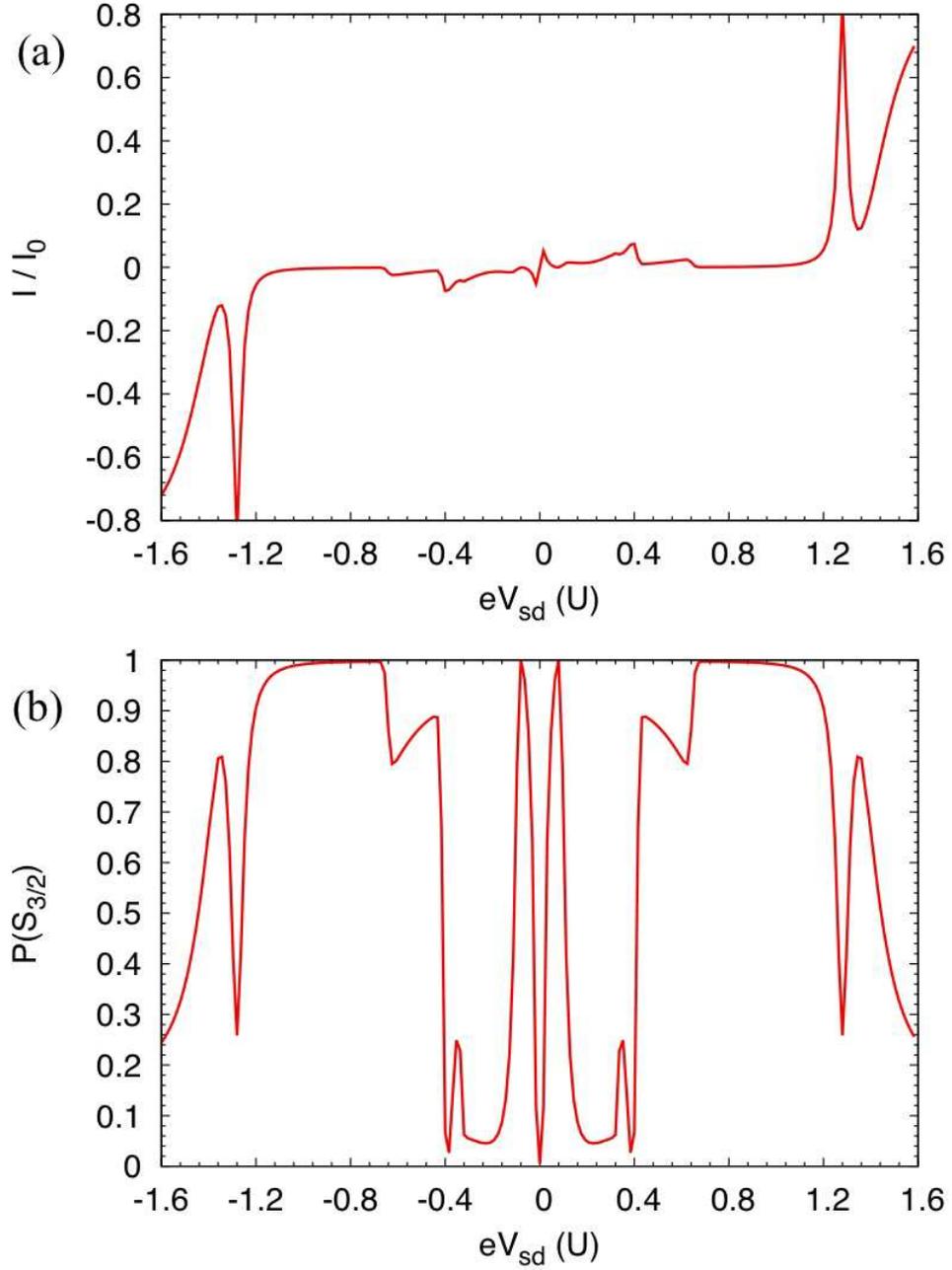}
\caption[I-V curve at symmetric quadruple point 2]{(Color Online) (a) Current of the LTQD at SQP as a function of $V_{sd}$.  At high bias (in both bias direction), we observe a robust negative differential conductance. (b) The steady-state occupation probability for spin-3/2 states. The peaks near the low- the figure  Indeed, the current suppression is associated with the spin-3/2 states.  The much more extended current suppression is due to the spin blockade phenomenon.}
\label{fig:sqp}
\end{figure}

\begin{figure}
\centering
\includegraphics[width=0.9\textwidth]{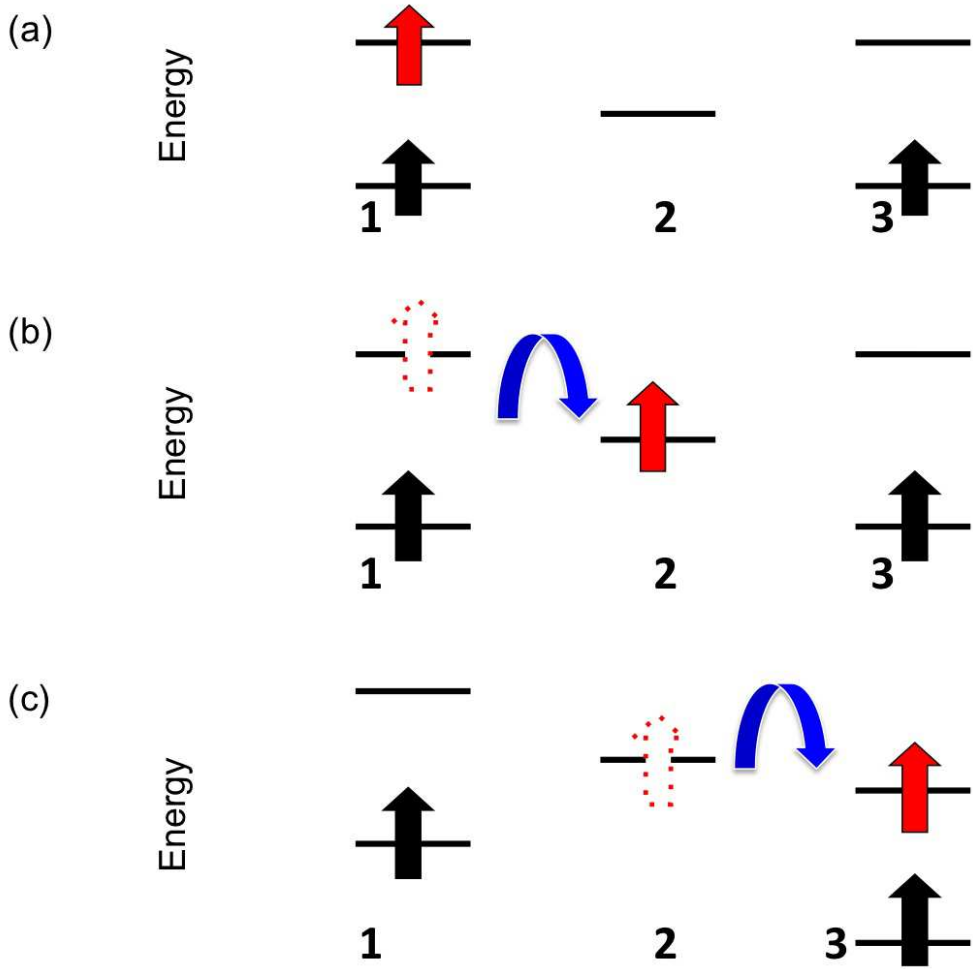}
\caption[Phonon-induced relaxation]{(Color Online) (Schematic representation of lifting of spin blockade. (a) (2,0,1) configuration obtained from the (1,0,1) configuration when an additional electron tunnels onto the $P$ orbital of dot 1.  (b) Due to phonon-induced relaxation in the model, the added spin moves from the $P$ orbital in dot 1 to the $S$ orbital in dot 2.  However, it does not proceed further to the dot 3 because this  costs energy.  (c) At larger source-drain bias in the positive direction, the energy levels in dot 3 are lowered with respect to that of dot 2.  Thus, phonon-induced relaxation assist the electron to move onto dot 2. }
\label{fig:phononblockade}
\end{figure}

\begin{figure}
\centering
\includegraphics[width=0.9\textwidth]{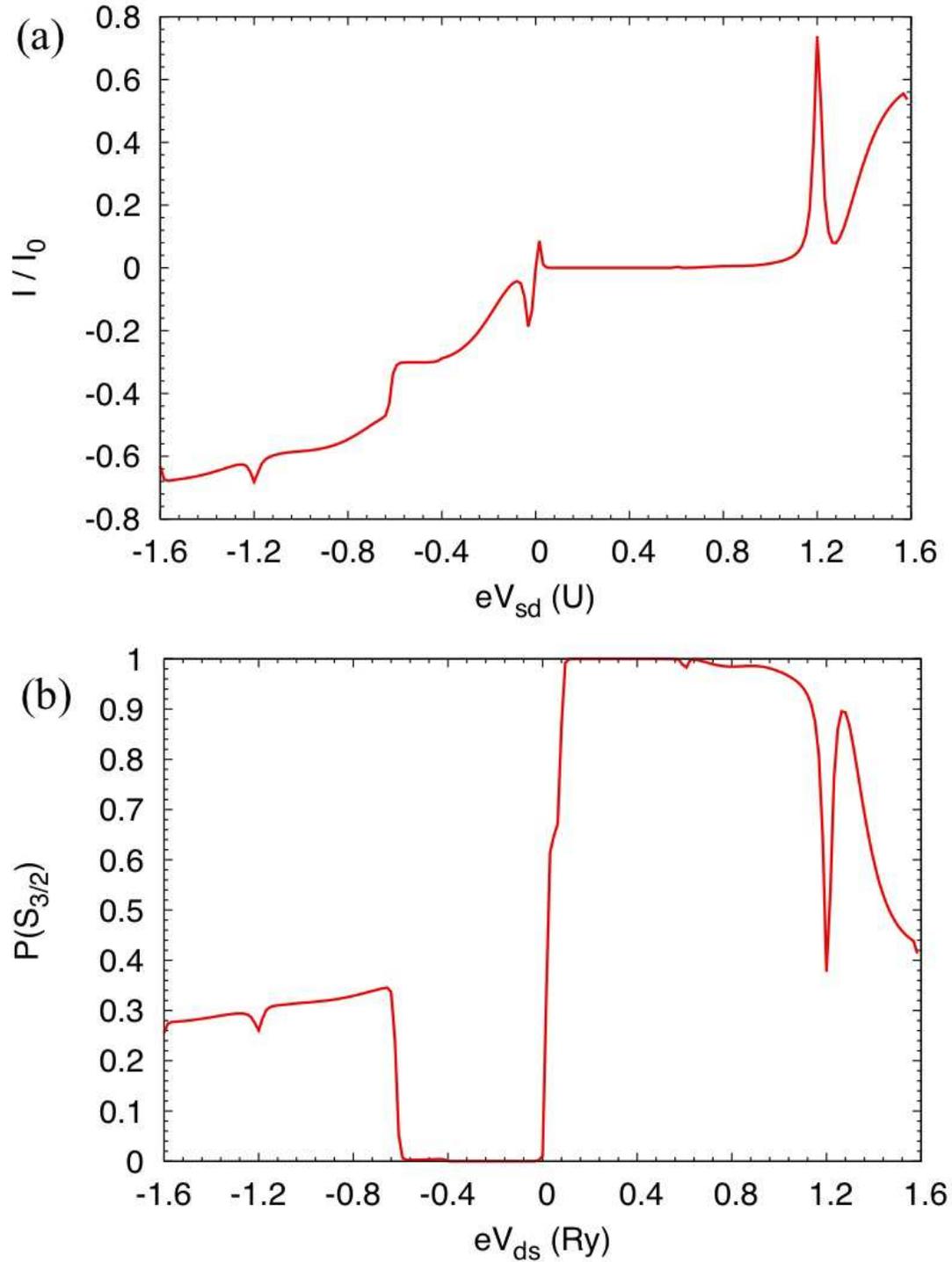}
\caption[I-V curve at asymmetric quadruple point]{(Color Online) (a) The current of the LTQD at AQP as a function of $V_{sd}$. The current response is asymmetrical with respect to the bias direction.  Similar to a DQD, current suppression is only observed in one direction of the bias.  (b) The steady-state occupation probability for spin-3/2 states. The current suppression is associated with the spin-3/2 states. }
\label{fig:aqp}
\end{figure}

\begin{figure}
\centering
\includegraphics[width=0.9\textwidth]{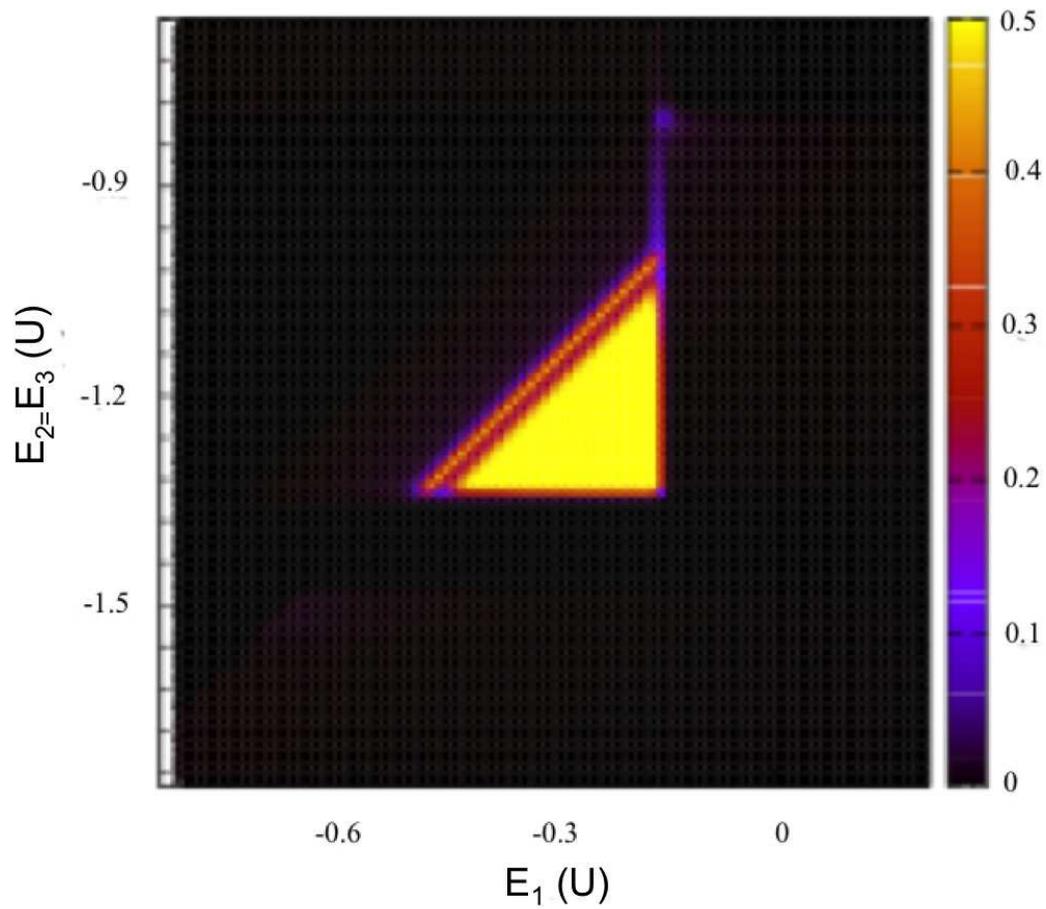}
\caption[Transport region in the on-site energy parameter space 2]{(Color Online) The current of the LTQD at AQP in the parameter space ($E_1, E_2=E_3$).}
\label{fig:aqp2}
\end{figure}

\end{document}